\documentclass[journal,10pt]{IEEEtran}

\ifCLASSINFOpdf
\else
  \usepackage[dvips]{graphicx}
\fi
\usepackage{url}
\usepackage[ruled,vlined]{algorithm2e}
\RestyleAlgo{ruled}
\usepackage{amsmath}
\usepackage{dsfont, cuted}
\usepackage{amsfonts}
\usepackage{amsthm}
\usepackage[utf8]{inputenc}
\usepackage{breqn}
\usepackage{xcolor}
\usepackage{subfigure}
\usepackage{multirow}
\usepackage{multicol}
\usepackage{booktabs}
\usepackage{fullwidth}
\usepackage[noadjust]{cite}
\usepackage{setspace}
\usepackage{hyperref}
\usepackage{bmpsize}
\usepackage{graphicx}
\usepackage{pbalance}
\usepackage{flushend}
\allowdisplaybreaks

\newtheorem{pro}{Proposition}
\newtheorem{thm}{Theorem}
\newtheorem{lem}{Lemma}

\newtheorem{rem}{Remark}
\newtheorem{prf}{Proof}

\newtheorem{defn}{Definition}

\begin{document}

\title{Robust Kalman Filters Based on the Sub-Gaussian $\alpha$-stable Distribution}

\author{Pengcheng~Hao, Oktay~Karakuş,~\IEEEmembership{Member,~IEEE,} Alin~Achim,~\IEEEmembership{Senior Member,~IEEE}
    \thanks{This work was supported by the Chinese Scholarship Council (CSC).}
    \thanks{Pengcheng Hao and Alin Achim are with the Visual Information Laboratory, University of Bristol, Bristol BS1 5DD, U.K. (e-mail: ju18422@bristol.ac.uk; alin.achim@bristol.ac.uk)}
    \thanks{Oktay Karakuş is with the School of Computer Science and Informatics, Cardiff University, Abacws, Cardiff, CF24 4AG, UK. (e-mail: karakuso@cardiff.ac.uk)}
}

\markboth{IEEE Transactions on XXXX, Vol. xx, No. x, 2023}
{Shell \MakeLowercase{\textit{et al.}}: Bare Demo of IEEEtran.cls for IEEE Journals}
\maketitle

\begin{abstract}

Motivated by filtering tasks under a linear system with non-Gaussian heavy-tailed noise, various robust Kalman filters (RKFs) based on different heavy-tailed distributions have been proposed. Although the sub-Gaussian $\alpha$-stable (SG$\alpha$S) distribution captures heavy tails well and is applicable in various scenarios, its potential has not yet been explored for RKFs. The main hindrance is that there is no closed-form expression of its mixing density. This paper proposes a novel RKF framework, RKF-SG$\alpha$S, where the signal noise is assumed to be Gaussian and the heavy-tailed measurement noise is modelled by the SG$\alpha$S distribution. The corresponding joint posterior distribution of the state vector and auxiliary random variables is approximated by the Variational Bayesian (VB) approach. Also, four different minimum mean square error (MMSE) estimators of the scale function are presented. The first two methods are based on the Importance Sampling (IS) and Gauss-Laguerre quadrature (GLQ), respectively. In contrast, the last two estimators combine a proposed Gamma series (GS) based method with the IS and GLQ estimators and hence are called GSIS and GSGL. Besides, the RKF-SG$\alpha$S is compared with the state-of-the-art RKFs under three kinds of heavy-tailed measurement noises, and the simulation results demonstrate its estimation accuracy and efficiency. {All the code needed to reproduce the results presented in this work are available at: https://github.com/PengchengH/Robust-Kalman-Filters-Based-on-the-Sub-Gaussian-alpha-stable-Distribution.}

\end{abstract} 


\IEEEpeerreviewmaketitle

\section{Introduction}

\IEEEPARstart{K}{alman} filtering is a technique that uses a sequence of observations to estimate the unknown states of a system over time~\cite{KF-1}. By judiciously defining the state vector of the relevant system, this algorithm can be employed and be beneficial in many applications. For example, in target tracking, the state vector represents the kinematics of the target~\cite{KF-2}; in GPS navigation, Kalman filter (KF) is applied to GPS receiver position/velocity determination~\cite{KF-3}; in robotics, the KF combines inertial, contact and kinematic data to estimate the robot’s base pose, velocity and contact states~\cite{KF-4}; in economics, it can capture interest rates, inflation, etc~\cite{KF-5}. When the dynamic and observation models of the state-space model are both linear and Gaussian, the minimum mean squared error (MMSE) estimate can optimally be obtained by the KF. However, when the noise contains outliers and is hence heavy-tailed, the estimation precision of the KF degrades since the Gaussianity assumption is violated.

To improve the performance of the KF under heavy-tailed noise, three main types of M-estimator-based robust KFs (RKFs) have been proposed. These are the Huber-based RKF (HKF)~\cite{HKF-1}, which is a generalized maximum likelihood-type estimator that minimises the Huber function, the maximum correntropy Kalman filter (MCKF)~\cite{MCKF-1}, and the statistical similarity measure based KF (SSMKF)~\cite{SSMKF-1}. {Alternatively, one has proposed several outlier-detection-based RKFs. For example,~\cite{RKF-outlier-Chisquare1,RKF-outlier-Chisquare2} detect the measurement outliers using the Chi-Square test, and variational Bayes strategies are employed in~\cite{RKF-outlier-VB1,RKF-outlier-VB2}.} However, these filters do not exploit the stochastic properties of the heavy-tailed noise, and hence their filtering performance is limited.

Different from M-estimator-based filters, another category of RKFs models the noise using heavy-tailed distributions. The Student's t (ST) distribution was the first model to be used for this purpose, and there were two types of approaches involving it. First, the ST-distribution-based RKFs assume that the measurements are contaminated by heavy-tailed noise. For example, in~\cite{RSTKF-1}, the symmetric measurement noise is modelled as an ST distribution. Also, skewed noise is fitted by the skewed ST distribution in~\cite{RSTKF-3}. However, when the signal noise is heavy-tailed, these filters are not applicable because of the Gaussian assumption which is violated. In a second group of techniques, RKFs are based on the assumption that both the one-step prediction and likelihood densities can be captured by ST distributions. In~\cite{RSTKF-4,RSTKF-5}, the ST-distribution-based filter (STF) considers that the heavy-tailed posterior probability density function (pdf) follows the ST distribution. However, its filtering accuracy is limited due to the requirement that the prediction and observation PDFs should be characterised by the same degree of impulsiveness. By contrast, in~\cite{RSTKF-6}, the robust Student’s t-based Kalman filter (RSTKF) eliminates this requirement and expresses the prediction and likelihood PDFs in hierarchical Gaussian forms. Hence, it exhibits higher computation complexity but produces more precise estimates. Despite all these improvements, two drawbacks of ST-distribution-based RKFs still remain. First, they provide a rough estimation when the noise is skewed. Also, the degree of freedom (dof) value of the ST distribution is set manually for each filtering scenario, which degrades its overall performance.

Owing to these disadvantages, other heavy-tailed distributions have also been investigated for the purpose of designing RKFs. In~\cite{RKF-LP-1,RKF-LP-2}, an RKF based on the multivariate Laplace distribution was considered. As the Laplace distribution is only determined by its mean and covariance, the selection of the dof value is unnecessary. Nevertheless, without a shape parameter, the filter cannot adapt to varying types of noise. Hence, for heavy-tailed and/or skewed state and measurement noises, the RKF based on the Gaussian scale mixture (RKF-GSM) distribution was proposed in~\cite{RKF-GSM-1}. However, the MMSE estimate of the scale function is not available, and the maximum \textit{a posteriori} probability (MAP) estimate introduces more errors. Also, the dof values are still manually set, which debases the estimation precision.

{This work focuses on linear models with heavy-tailed sub-Gaussian $\alpha$-stable (SG$\alpha$S) noise, and various RKFs can be used for the posterior estimation. However, the M-estimator-based and outlier-detection-based RKFs cannot exploit the stochastic characteristics of the SG$\alpha$S noise, and the performance of the RKFs using heavy-tailed distributions might degrade due to the model error between the utilised heavy-tailed distributions and the SG$\alpha$S noise. Hence, a new RKF framework based on the SG$\alpha$S} distribution~\cite{SGS-1,SGS-2} is considered in this work. The motivation for this choice is twofold. 1) The SG$\alpha$S distribution belongs to the $\alpha$-stable family of distributions and is hence theoretically motivated by the generalized Central Limit Theorem (GCLT), which means many practical types of noise can be modelled by this density~\cite{SGS-application-1,SGS-application-2}.
{For example, work by Lachmann in the early 20th century~\cite{APP-1} shows that the SG$\alpha$S distribution can well fit the noise in real audio data and filtering based on a SG$\alpha$S distribution was proposed for music enhancement. Also,~\cite{APP-2} and~\cite{APP-3} utilise the SG$\alpha$S distribution to inflation forecasting in macroeconomic analysis and acoustic source target tracking, respectively, where the isotropic complex symmetric $\alpha$ stable (S$\alpha$S) noise can be considered a SG$\alpha$S variable~\cite{SGS-1}. Besides, the multiple access interference (MAI) in ad hoc networks can be precisely represented by the SG$\alpha$S distribution~\cite{APP-4}.} {2) Besides being a member of the more general family of alpha-stable distributions, the SG$\alpha$S is also a special case of the GSM class of distributions~\cite{SGS-3,SGS-4}. Hence, the design of an RKF framework based on the SG$\alpha$S distribution can, in principle, be based on the structure of the RKF-GSM. However, as there is no closed-form expression for its mixing density, the MMSE estimation of the scale function cannot be achieved simply based on the RKF-GSM and hence remains challenging. Thus, despite its many appealing properties and various practical applications, the use of the SG$\alpha$S distribution for the design of RKF filters has not been investigated, and this constitutes the primary focus of this manuscript.}

Specifically, the main contributions of this work consist in:
\begin{enumerate}
  \item  {We propose a RKF framework based on the SG$\alpha$S distribution (RKF-SG$\alpha$S), where four different estimators are proposed to approximate the MMSE of the scale function. For the first two methods, the expectation of the scale function is approximated by the Importance Sampling (IS)~\cite{IS-1} and Gauss-Laguerre quadrature (GLQ) methods~\cite{GLQ-1}, respectively. Two additional hybrid estimators, the gamma series IS (GSIS) and gamma series GLQ (GSGL), are proposed for situations when the proposed GS estimator diverges and is thus replaced with the IS and GLQ-based methods, respectively.}
  
  \item The properties of the four proposed RKF-SG$\alpha$S variants based on the different estimators are thoroughly analysed and the RKF-SG$\alpha$S is compared with the benchmark heavy-tailed-distribution-based RKFs in different measurement noise environments. Particularly, the parameters of the SG$\alpha$S distribution {and the benchmark GSM distributions are estimated based on the existing expectation maximisation (EM) and maximum likelihood estimation (MLE) methods, respectively, which reduces the model errors caused by the empirical parameters.} 
\end{enumerate}
As a side contribution, for the RKF based on the slash distribution (RKF-SL)\cite{RKF-GSM-1}, which is one of the filters we benchmark against in the simulations section, we propose a novel MMSE estimate of the scale function in Appendix~\ref{sec:RKF-SL}. 

The rest of the paper is organised as follows: we begin, in Section~\ref{sec:Theoretical preliminary}, with a description of the theoretical background on the Linear Discrete-Time State-Space Model and the SG$\alpha$S distribution. In section~\ref{sec:RKF-SSG}, we present the details of the proposed RKF-SG$\alpha$S. The proposed framework is tested in target tracking scenarios in Section~\ref{sec: experiments}, whilst Section~\ref{sec:conclusion} concludes the paper with a summary.

\section{Theoretical preliminary} \label{sec:Theoretical preliminary}
In this section, we provide brief, necessary details, on the key concepts on which our main developments in subsequent sections build.
\subsection{Nomenclature}
\begin{table} [h!]
\centering
\begin{tabular}{p{1.75cm} p{6.25cm}}\toprule
\textit{Notations}    & \textit{Definitions}  \\ \toprule
$\mathcal{N}(\boldsymbol{\mu},\mathbf{\mathbf{\Sigma}})$, 
$\mathcal{N}(.;\boldsymbol{\mu},\mathbf{\mathbf{\Sigma}})$ 
             & Multivariate Gaussian pdf with mean vector $\boldsymbol{\mu}$ and covariance matrix $\mathbf{\mathbf{\Sigma}}$. \\\hline
$S(.;\alpha,\beta,\gamma,\delta)$      & Univariate stable distribution pdf with the shape parameter $\alpha$, the skewness parameter $\beta$, the scale parameter $\gamma$ and the location parameter $\delta$.\\\hline
$\mathrm{G}(.;a,b)$    & Gamma pdf with the shape parameter $a$ and rate parameter $b$. \\\hline
$\mathrm{Be}(.;a,b)$   & Beta pdf with the shape parameter $a$ and scale parameter $b$.  \\\hline
$\mathrm{IG}(.;a,b)$   & Inverse-Gamma pdf with the shape parameter $a$ and $b$.  \\ \hline
$\mathrm{IW}(.;v,\mathbf{\mathbf{\Sigma}})$ 
             & Inverse-Wishart pdf with the dof parameter $v$ and scale matrix $\mathbf{\mathbf{\Sigma}}$. \\\hline
$\Gamma(\cdot)$, $\psi(\cdot)$        & Gamma function and Digamma function. \\\hline
$\mathit{\gamma}(a,x)$       & Lower Incomplete Gamma function defined as $\int_0^xt^{a-1}e^{-t}dt$. \\\hline
$\mathbf{I}_n$      & $n \times n$ identity matrix. \\\hline
$\mathrm{tr}\{\}$,$(\cdot)^T$ & Trace operation and transpose operation.\\\hline
$|\cdot|$                  & Absolute value operation. \\\hline
$\mathbf{abs}(\mathbf{X})$ &  Element-wise absolute value operation. \\\hline
$\mathbf{sum}(\mathbf{X})$ &  Summation for all the elements. \\\hline
$\mathbf{x}\sim p(\mathbf{x})$ & The pdf of the variable $\mathbf{x}$ is $p(\mathbf{x})$. \\\hline
$\stackrel{d}{=}$, $\stackrel{d}{\rightarrow}$ & Equality and convergence in distribution. \\\bottomrule
\end{tabular}
\label{table:notation}
\end{table}

\subsection{Linear Discrete-Time State-Space Model}
A linear discrete-time state-space model can be described as a probabilistic graph structure and includes two models. The signal model describes the state transition over time, whilst the measurement model explains the relationship between the states and corresponding measurements. A generic linear discrete-time state-space model is written as
\begin{equation} \label{eq:statespace_1}
\begin{cases}  \mathbf{x}_k=\mathbf{F}_k\mathbf{x}_{k-1}+\mathbf{w}_{k-1}\\ \mathbf{z}_k=\mathbf{H}_k\mathbf{x}_k+\mathbf{v}_k
\end{cases}
\end{equation}
where $\mathbf{x}_k\in \mathbb{R}^n$ and $\mathbf{z}_k\in \mathbb{R}^m$ denote the hidden state and the measurement vectors at time $k$, respectively. Also the process and measurement noises are represented as $\mathbf{w}_{k}\in \mathbb{R}^n$ and $\mathbf{v}_k\in \mathbb{R}^m$, respectively. Additionally, $\mathbf{F}_k\in\mathbb{R}^{n\times n}$ and $\mathbf{H}_k\in \mathbb{R}^{m\times n}$ are the state transition and measurement matrices. In this work, we assume $\mathbf{w}_{k}$ follows a Gaussian distribution, whilst $\mathbf{v}_{k}$ contains outliers and is thus heavy-tailed.

\subsection{$\alpha$-stable Distributions and the Sub-Gaussian Case} \label{sec:SSG}
In this section, we first introduce the $\alpha$-stable family of distributions whereby its sub-Gaussian subclass, the SG$\alpha$S distribution, is also illustrated. Furthermore, the tail behaviour of this subclass is analysed.

In probability theory, the $\alpha$-stable distribution is a family of probability distributions which is generally used to model heavy-tailed behaviour and outlier effects~\cite{SGS-model-1,SGS-model-2,SGS-application-1}. Particularly, a univariate stable random variable can be defined based on the following stability property~\cite{SGS-3}:
\begin{defn} \label{def:stability}
A random variable $x$, has an $\alpha$-stable distribution if for any positive numbers $A$ and $B$, there is a positive number $C$ and a real number $D$ such that
\begin{equation*}
    Ax_1+Bx_2\stackrel{d}{=}Cx+D
\end{equation*}
where $x_1$ and $x_2$ are independent copies of $x$.
\end{defn}
Also, a stable random variable can be defined based on the generalized Central Limit Theorem~\cite{SGS-3}:
\begin{defn} \label{def:GCLT}
A random variable $x$ has a stable distribution if it has a domain of attraction, i.e., if there is a sequence of i.i.d. random variables $y_1, y_2, \dots$ and sequence of positive numbers $(d_{n^\prime})_{(n^\prime \in \mathbf{N})}$ and real numbers $(D_{n^\prime})_{(n^\prime \in \mathbf{N})}$, such that
\begin{equation*}
\frac{y_1+y_2+\dots+y_{n^\prime}}{d_{n^\prime}}+D_{n^\prime} \stackrel{d}{\rightarrow} x
\end{equation*}
\end{defn} 
The next definition represents a stable variable in terms of its characteristic function~\cite{SGS-1}.
\begin{defn} \label{def: CF of univariate}
A random variable is stable if its characteristic function can be written as
\begin{equation*}
\begin{split}
&\mathrm{E}\left(\mathrm{exp}(\mathtt{i}\theta x)\right)= \\    
&
\begin{cases}
 \mathrm{exp}\left(-\gamma^\alpha|\theta|^\alpha\left[1-\mathtt{i}\beta\left(\mathrm{tan}\frac{\pi \alpha}{2}\right)\left(\mathrm{sign}\,\theta\right)\right]+\mathrm{i}\delta\theta\right), \alpha \neq 1 \\
 \mathrm{exp}\left(-\gamma|\theta|\left[1+\mathtt{i}\beta\frac{2}{\pi}\left(\mathrm{sign}\,\theta\right) \mathrm{ln} |\theta|\right]+\mathrm{i}\delta\theta\right)\qquad\, ,\alpha = 1
\end{cases}
\end{split}
\end{equation*}
where the stability parameter $\alpha \in (0,2]$, skewness parameter $\beta \in [-1,1]$, scale parameter $\gamma \in (0,\infty)$ and location parameter $\delta \in \mathbb{R}$.    
\end{defn}
Furthermore, the multivariate $\alpha$-stable distribution extends the concept of the univariate stable distribution to high-dimensional cases. Thus, stable random vectors also exhibit the stability property and allow for the GCLT. However, we cannot directly apply the multivariate stable distribution to RKF frameworks due to the challenge of estimating the spectral measure of its characteristic function~\cite{SGS-3}. Instead, we focus on a more suitable subclass of the multivariate stable distribution, known as the SG$\alpha$S distribution. It has a simpler spectral measure, and its characteristic function can be expressed as~\cite{SGS-4}
\begin{equation*}
\mathrm{E}\left[\mathrm{exp}(\mathrm{i}\boldsymbol{\theta}^T\mathbf{x})\right] =\mathrm{exp}\left[\mathrm{i}\boldsymbol{\theta}^T\boldsymbol{\mu} - \left(\boldsymbol{\theta}^T\Sigma\boldsymbol{\theta}\right)^{\frac{\alpha}{2}}\right].
\end{equation*}
As previously noted, the SG$\alpha$S distribution is also a special case of 
the GSM distribution which can be written as
\begin{equation} \label{eq:GSM}
p(\mathbf{x})= \int_0^{+\infty} \mathcal{N}(\mathbf{x}; \boldsymbol{\mu}+y\boldsymbol{\beta},\mathbf{\mathbf{\Sigma}}/\kappa(y))\pi(y)dy,
\end{equation}
where $\boldsymbol{\mu}$ is the mean vector and $\boldsymbol{\beta}$ is the skewness vector. Also, $\mathbf{\Sigma}$ is the scale matrix, $y>0$ is the mixing parameter, and $\kappa(y)$ and $\pi(y)$ are the scale function and mixing density, respectively. For the SG$\alpha$S distribution~\cite{SGS-3}, $\boldsymbol{\beta}=\mathbf{0}$, $\kappa(y)=1/y$ and $\pi(y)$ can be represented as a totally skewed univariate stable distribution, i.e.,
\begin{equation} \label{eq: SSG}
p(\mathbf{x})= \int_0^{+\infty} \mathcal{N}(\mathbf{x}; \boldsymbol{\mu},y\mathbf{\mathbf{\Sigma}})S(y;\alpha/2,1,\textrm{cos}(\pi\alpha/2)^{2/\alpha},0)dy.    
\end{equation}
Table~\ref{table:GSM} details the parameters of SG$\alpha$S along with several GSM distributions investigated in this paper, where $v$ refers to the dof parameter. For simplicity, we only consider the zero-mean symmetric heavy-tailed noise as in previous works~\cite{RSTKF-1,RKF-GSM-2} where the mean and skewness vectors are zero for both the SG$\alpha$S and the GSM distributions. 

\begin{table*}
\centering
\caption{EXEMPLARY GSM DISTRIBUTIONS AND THEIR PARAMETERS}
\begin{tabular}{c c c c}
\hline
\textbf{GSM distribution} & \textbf{Scale function} & \textbf{Mixing density} & \textbf{Constraints}  \\ 
SG$\alpha$S & $\kappa(y)=y^{-1}$ & $\pi(y)=S(y;\alpha/2,1,\textrm{cos}(\pi\alpha/2)^{2/\alpha},0)$ & $y>0,0<\alpha\leq2$ \\
Student's $t$  & $\kappa(y)=y$  & $\pi(y)=\mathrm{G}(y;\frac{v}{2},\frac{v}{2})$ & $y>0,v>0$\\
Slash    & $\kappa(y)=y$  & $\pi(y)=\mathrm{Be}(y;\frac{v}{2},1)$ & $0<y<1,v>0$\\
Variance Gamma & $\kappa(y)=y$  & $\pi(y)=\mathrm{IG}(y;\frac{v}{2},\frac{v}{2})$ & $y>0,v>0$ \\ 
\hline
\end{tabular}
\label{table:GSM}
\end{table*}

\begin{figure}[h!]
\centering
\includegraphics[width=\linewidth]{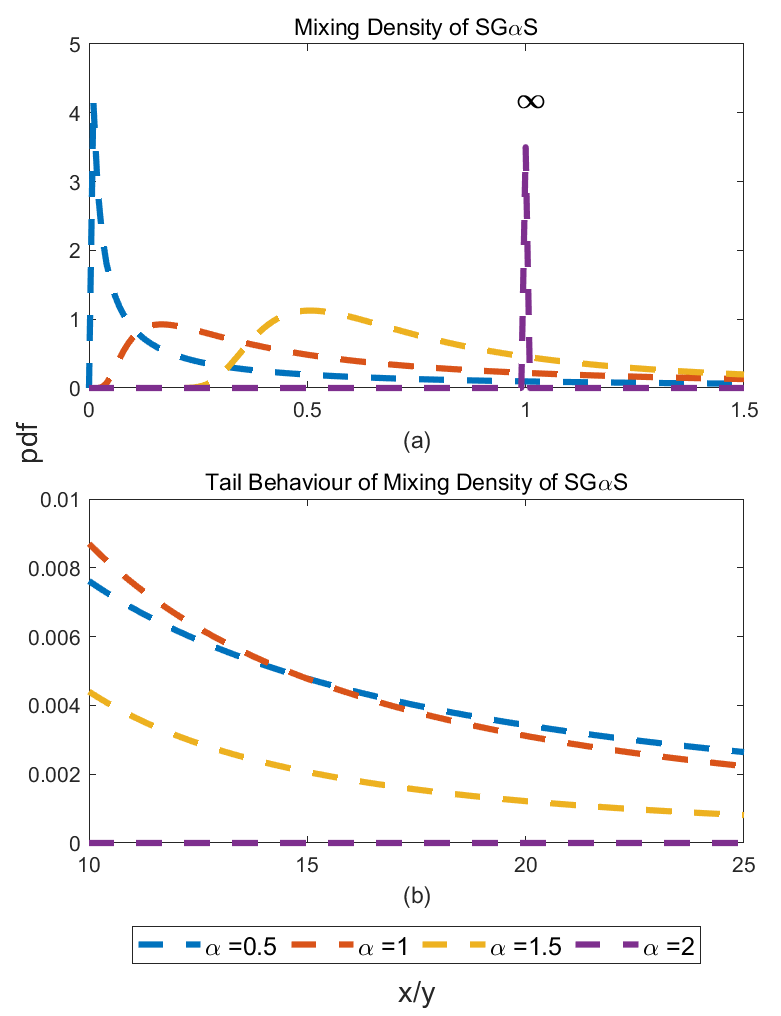}
\caption{Mixing density and tail behaviour analysis. (a) and (b) plot the pdf and tail behaviours of the mixing density $S(y;\alpha/2,1,\textrm{cos}(\pi\alpha/2)^{2/\alpha},0)$, respectively.}
\label{fig: stable density}
\end{figure}

Subsequent to the succint introduction of the SG$\alpha$S distribution, we explain its tails behaviour, which is determined by its mixing density. Figure~\ref{fig: stable density}-(a) shows the mixing density is a Dirac delta distribution at $1$ for $\alpha=2$ and a totally skewed heavy-tailed stable distribution for $\alpha<2$. Besides, smaller $\alpha$ values result in heavier tails as in Figure~\ref{fig: stable density}-(b).

\section{RKF-SG$\alpha$S} \label{sec:RKF-SSG}
This section begins by introducing the proposed robust Kalman filtering framework based on the SG$\alpha$S distribution in section~\ref{sec:RKF-framework}. Subsequently, Section~\ref{sec:scale-estimator} details the four MMSE estimators used to estimate the scale function. Moreover, Section~\ref{sec:convergence-test} describes the convergence test strategy employed for the RKF-SG$\alpha$S.

\subsection{Proposed Robust Kalman Filtering Framework} \label{sec:RKF-framework}
Assume the zero-mean signal and measurement noises are subject to Gaussian and SG$\alpha$S distributions, respectively. Then, both are modelled as 
\begin{align}\label{eq:noise model}
p(\mathbf{w}_{k-1})&= \mathcal{N}(\mathbf{0},\mathbf{Q}_{k-1}) \\
p(\mathbf{v}_{k})&= \int_0^{+\infty} \mathcal{N}(\mathbf{v}_k;\mathbf{0},\lambda_k\mathbf{R}_k)S(\lambda_k) d\lambda_k 
\end{align}
where $\mathbf{Q}_{k-1}$ is the covariance matrix of the state noise at time $k-1$ whilst $\mathbf{R}_{k}$ and $\lambda_k$ refer to the scale matrix and mixing parameter of the measurement noise at time $k$, respectively. Also, the mixing density $S(\lambda_k;\alpha/2,1,\textrm{cos}(\pi\alpha/2)^{2/\alpha},0)$ is simplified as $S(\lambda_k)$. Hence, the forecast pdf $p(\mathbf{x}_k|\mathbf{z}_{1:k-1})$ and likelihood pdf $p(\mathbf{z}_{k}|\mathbf{x}_k)$ can be expressed as
\begin{align}
 p(\mathbf{x}_k|\mathbf{z}_{1:k-1})&=\mathcal{N}(\mathbf{x}_k; \mathbf{F}_k\mathbf{\hat{x}}_{k-1|k-1},\mathbf{P}_{k|k-1}) \label{eq:forcast pdf} \\ 
 p(\mathbf{z}_{k}|\mathbf{x}_k)&=\int_0^{+\infty}\mathcal{N}(\mathbf{z}_k; \mathbf{H}_k\mathbf{x}_{k},\lambda_k\mathbf{R}_k)S(\lambda_k)d\lambda_k \label{eq: likelihood pdf}
\end{align}
where $\mathbf{\hat{x}}_{k-1|k-1}$ is the posterior mean vector at time $k-1$ and 
\begin{equation} \label{eq:Pk_k-1}
 \mathbf{P}_{k|k-1}=\mathbf{F}_k\mathbf{P}_{k-1|k-1}\mathbf{F}_k^T+\mathbf{Q}_{k-1}   
\end{equation}
denotes the error covariance matrix at time $k-1$. According to~(\ref{eq: likelihood pdf}), $p(\mathbf{z}_{k}|\mathbf{x}_k)$ follows a hierarchical Gaussian form of
\begin{equation} \label{eq:hierarchical Gaussian}
p(\mathbf{z}_{k}|\mathbf{x}_k,\lambda_k,\mathbf{R}_k)=\mathcal{N}(\mathbf{z}_k; \mathbf{H}_k\mathbf{x}_{k},\lambda_k\mathbf{R}_k)
\end{equation}
where $\lambda_k\sim S(\lambda_k)$. Assume the true value of $\mathbf{R}_k$ is unknown, and we model its uncertainty using the inverse-Wishart (IW) distribution, i.e. $p(\mathbf{R}_k)= \mathrm{IW}(\mathbf{R}_k;u_k,\mathbf{U}_k)$, where $u_k$ and $\mathbf{U}_k$ are the dof parameter and scale matrix, respectively. According to equations~(\ref{eq:noise model})-(\ref{eq:hierarchical Gaussian}), the joint posterior distribution $p(\boldsymbol{\Theta}|\mathbf{z}_{1:k})$ can be expressed as
\begin{equation}
\begin{split}
p(\boldsymbol{\Theta}|\mathbf{z}_{1:k})&\propto p(\mathbf{z}_{k}|\boldsymbol{\Theta})p(\boldsymbol{\Theta}|\mathbf{z}_{1:k-1})p(\mathbf{z}_{1:k-1}) \\
 &=\mathcal{N}(\mathbf{z}_k; \mathbf{H}_k\mathbf{x}_{k},\lambda_k\mathbf{R}_k) \\
 &\times \mathcal{N}(\mathbf{x}_k; \mathbf{F}_k\mathbf{\hat{x}}_{k-1|k-1},\mathbf{P}_{k|k-1})\times S(\lambda_k) \\
 &\times \mathrm{IW}(\mathbf{R}_k;u_k,\mathbf{U}_k)\times p(\mathbf{z}_{1:k-1}) 
\end{split}
\end{equation}
where $\boldsymbol{\Theta}_k= \{\mathbf{x}_k,\lambda_k,\mathbf{R}_k\}$.
As there is no analytic expression for $p(\boldsymbol{\Theta}_k|\mathbf{z}_{1:k})$, the Variational Bayesian (VB) approach~\cite{VB-1} based on the EM algorithm is employed to approximate the posterior density from
\begin{equation}
p(\boldsymbol{\Theta}_k|\mathbf{z}_{1:k}) \approx q(\mathbf{x}_k)q(\lambda_k)q(\mathbf{R}_k).
\end{equation}
Then, the density $q(\cdot)$ can be calculated by
\begin{equation}\label{eq:VB}
\mathrm{log}~q(\boldsymbol{\phi})=\mathrm{E}_{\boldsymbol{\Theta}_k^{(-\phi)}} \left[\mathrm{log}~ 
p(\boldsymbol{\Theta}_k,\mathbf{z}_{1:k}) \right]+c_{\phi}
\end{equation}
where $\boldsymbol{\phi}\subset\boldsymbol{\Theta}_k$, $\boldsymbol{\phi}\cup\boldsymbol{\Theta}_k^{(-\phi)}=\boldsymbol{\Theta}_k$ and $c_{\phi}$ is a constant number related to $\phi$. 

Next, the fixed-point iteration is employed to solve equation~(\ref{eq:VB}). At the $t+1$-th iteration, $q^{(t+1)}(\boldsymbol{\phi})$ is approximated by employing $q^{(t)}(\boldsymbol{\phi})$ to calculate the expectation in~(\ref{eq:VB}). Proposition~\ref{pro:xk}-\ref{pro:Rk} present the updated formulation, but their proofs are not provided in this work as similar proofs have been made in previous work~\cite{RKF-GSM-1,RKF-GSM-2,RKF-GSM-3}.
\begin{pro}[\cite{RKF-GSM-2} Proposition 2]\label{pro:xk}
Let $\boldsymbol{\phi}=\mathbf{x}_k$, then $q^{(t)}(\mathbf{x}_k)$ can be updated as the Gaussian distribution, i.e.,
\begin{equation*}
 q^{(t+1)}(\mathbf{x}_k)=\mathcal{N}\left(\mathbf{x}_k;\mathbf{\hat{x}}_{k|k}^{(t+1)},\mathbf{P}_{k|k}^{(t+1)}\right),
\end{equation*}
where
\begin{equation}\label{eq:KF update}
\begin{split}
\mathbf{\hat{x}}_{k|k-1}&=\mathbf{F}_k\mathbf{\hat{x}}_{k-1|k-1} \\ 
\mathbf{K}_{k}^{(t+1)}&=\mathbf{P}_{k|k-1} \mathbf{H}_k^T\left(\mathbf{H}_k\mathbf{P}_{k|k-1}^T\mathbf{H}_k^T+\mathbf{\Tilde{R}}_k^{(t)}\right)^{-1}\\
\mathbf{\hat{x}}_{k|k}^{(t+1)}&=\mathbf{\hat{x}}_{k|k-1}+\mathbf{K}_{k}^{(t+1)}\left(\mathbf{z}_k-\mathbf{H}_k\mathbf{\hat{x}}_{k|k-1}\right) \\
\mathbf{P}_{k|k}^{(t+1)}&=\left(\mathbf{I}_n-\mathbf{K}_{k}^{(t+1)}\mathbf{H}_k\right)\mathbf{P}_{k|k-1}.
\end{split}
\end{equation}
where $\mathbf{\Tilde{R}}_k^{(t)}$ is the modified measurement noise covariance matrix and can be written as
\begin{equation} \label{eq:modified covariance}
\mathbf{\Tilde{R}}_k^{(t)}=\frac{ \left[\mathrm{E}^{(t)}\left(\mathbf{R}_k^{-1}\right) \right]^{-1} }{\mathrm{E}^{(t)}\left(\kappa(\lambda_k)\right)}
=\frac{ \left[\mathrm{E}^{(t)}\left(\mathbf{R}_k^{-1}\right) \right]^{-1} }{\mathrm{E}^{(t)}\left(\lambda_k^{-1}\right)}. 
\end{equation}
\end{pro}

\begin{pro}[\cite{RKF-GSM-2} Proposition 4] \label{pro:lambdak}
Let $\boldsymbol{\phi}=\lambda_k$, $\mathrm{\log}q^{(t+1)}(\lambda_k)$ can be expressed as,
\begin{equation} \label{eq:q_lambda}
 \mathrm{\log}q^{(t+1)}(\lambda_k)=-\frac{\eta}{2}\lambda_k^{-1}-\frac{m}{2}\mathrm{\log}\lambda_k+\mathrm{\log}S(\lambda_k)+C^\prime,
\end{equation}
where $C^\prime$ is a constant number and
\begin{align}
\begin{split} \label{eq:eta and B}
\eta&=\mathrm{tr}\left\{\mathbf{B}_k^{(t+1)}\mathrm{E}^{(t)}\left(\mathbf{R}_k^{-1}\right)\right\} \\
\mathbf{B}_k^{(t+1)}
&=\left(\mathbf{z}_k-\mathbf{H}_k\mathbf{\hat{x}}_{k|k}^{(t+1)}\right)\left(\mathbf{z}_k-\mathbf{H}_k\mathbf{\hat{x}}_{k|k}^{(t+1)}\right)^T \\
&\quad+\mathbf{H}_k\mathbf{P}_{k|k}^{(t+1)}\mathbf{H}_k^T
\end{split}
\end{align}
\end{pro}

\begin{pro}[\cite{RKF-GSM-3} Proposition 4] \label{pro:Rk}
 Let $\boldsymbol{\phi}=\mathbf{R}_k$, $q^{(t+1)}(\mathbf{R}_k)$ is written as the IW distribution, i.e.,
 \begin{equation*}
 q^{(t+1)}(\mathbf{R}_k)=\mathrm{IW}\left(\mathbf{R}_k;u_k^{(t+1)},\mathbf{U}_k^{(t+1)}\right),
 \end{equation*}
 where 
\begin{equation}\label{eq:parameters of IW}
\begin{split}
     u_k^{(t+1)}&=u_k+1, \mathbf{U}_k^{(t+1)}=\mathbf{U}_k+\mathbf{D}_k^{(t+1)} \\
    \mathbf{D}_k^{(t+1)}&=\mathrm{E}^{(t+1)}\left(\lambda_k^{-1}\right) \\
    &\times \left[\mathbf{b}_k^{(t+1)}\left(\mathbf{b}_k^{(t+1)}\right)^T+\mathbf{H}_k\mathbf{\hat{P}}_{k|k}^{(t+1)}\mathbf{H}_k^T\right] \\
    \mathbf{b}_k^{(t+1)}&=\mathbf{z}_k-\mathbf{H}_k\mathbf{\hat{x}}_{k|k}^{(t+1)}   
\end{split}
\end{equation}
Thence, we have
\begin{equation}\label{E_Rk}
\mathrm{E}^{(t+1)}\left(\mathbf{R}_k^{-1}\right)=\left(u_k^{(t+1)}-m-1\right)\left(\mathbf{U}_k^{(t+1)}\right)^{-1}
\end{equation}
\end{pro}

\begin{rem}
In equations~(\ref{eq:modified covariance}) and~(\ref{eq:parameters of IW}), the expectation of the scale function $\mathrm{E}\left(\kappa(\lambda_k)\right)=\mathrm{E}\left(\lambda_k^{-1}\right)$ is required. While Proposition~\ref{pro:lambdak} provides the pdf of $\lambda_k$, the calculation of the expectation is not straightforward due to the lack of a closed-form expression of $S(\lambda_k)$ in equation~(\ref{eq:q_lambda}).
\end{rem}

\subsection{MMSE Estimators of the Scale Function}\label{sec:scale-estimator}
To estimate the expectation of the scale function, this section introduces four estimators of $\mathrm{E}^{(t+1)}\left(\lambda_k^{-1}\right)$, including IS and GLQ-based estimators, along with their corresponding hybrid estimators with the incorporation of the GS based estimator. For simplicity, we replace $\lambda_k$ with $y$ and omit the iteration number $t+1$ in~(\ref{eq:q_lambda}), and then we have 
\begin{align}
&\mathrm{E}^{(t+1)}\left(\lambda_k^{-1}\right) =\mathrm{E}\left(y^{-1}\right) \\
&\mathrm{log}q(y)=-\frac{\eta}{2}y^{-1}-\frac{m}{2}\mathrm{log}y+\mathrm{log}S(y)+C^\prime \label{eq:logq_y}\\
&q(y) \propto q^ \prime (y)= y^{-\frac{m}{2}}\mathrm{exp}\left(-\frac{\eta}{2y}\right)S(y), \label{eq:q_y} 
\end{align} 
where $q^ \prime (y)$ is a function proportional to the density $q(y)$. Then,
\begin{equation} \label{eq:Ey^{-1}}
\mathrm{E}\left(y^{-1}\right)=\frac{\int_0^{+\infty}y^{-1}q^ \prime(y)dy}{\int_0^{+\infty}q^ \prime(y)dy}.
\end{equation}

\subsubsection{Importance Sampling Based Estimator}
Here we introduce the estimation method of $\mathrm{E}\left(y^{-1}\right)$ based on the IS technique. The IS-based estimator is elaborated in Theorem~\ref{thm:IS}.
\begin{thm} \label{thm:IS}
Let $y_i~(i=1,\dots,N)$ be samples from the target distribution $S(y)$. Then $\mathrm{E}\left(y^{-1}\right)$ can be estimated by
\begin{equation} \label{eq:IS-1}
\mathrm{E}\left(y^{-1}\right)=\sum_{i=1}^Nw_iy_i^{-1}  
\end{equation}
where
\begin{equation} \label{eq:IS-2}
w_i=\frac{y_i^{-\frac{m}{2}}\mathrm{exp}\left(-\frac{\eta}{2y_i}\right)}{\sum_{i^\prime=1}^Ny_{i^\prime}^{-\frac{m}{2}}\mathrm{exp}\left(-\frac{\eta}{2y_{i^\prime}}\right)}.  
\end{equation}
\end{thm}
\textit{Proof}: See Appendix~\ref{proof:thm:IS} \qedsymbol.
\begin{rem}
Although the approximation of the  $S(y)$ value in~(\ref{eq:q_y}) is computationally expensive~\cite{SGS-model-6}, it can be efficiently sampled~\cite{SGS-model-4}. Hence, based on the IS algorithm, we first obtain the samples of $S(y)$, and then the numerator and denominator integrals in~(\ref{eq:Ey^{-1}}) can be approximated.   
\end{rem}

\subsubsection{Gauss–Laguerre Quadrature Based Estimator}
An alternative estimator of $\mathrm{E}\left(y^{-1}\right)$, this time based on the GLQ, is discussed in this section. The details of this estimator are provided in Theorem~\ref{thm:GL}.
\begin{thm} \label{thm:GL}
Based on the GLQ method, $\mathrm{E}\left(y^{-1}\right)$ can be approximated by
\begin{equation} \label{eq:GL-1}
\mathrm{E}\left(y^{-1}\right)=\frac{\sum_{l=1}^Lw_lx_lf(x_l)}{\frac{\eta}{2} \sum_{l=1}^Lw_lf(x_l)}.
\end{equation}
where $x_l$, $l=1,\dots,L$, is the $l$-th root of Laguerre polynomial $\mathcal{L}_L(x)$ and $f(x_l)=x_l^{\frac{m}{2}-2}S(\frac{\eta}{2x_l})$.
Also, the weight $w_l$ can be written as
\begin{equation}\label{eq:GL-2}
w_l=\frac{x_l}{(L+1)^2\left[\mathcal{L}_{L+1}(x_l)\right]^2}.
\end{equation}
\end{thm}
\textit{Proof}: See Appendix~\ref{proof:thm:GL}.
\begin{rem}
The GLQ-based estimator in Theorem~\ref{thm:GL} requires calculating $f(x_l)$ in which the approximation of $S(\frac{\eta}{2x_l})$ cannot be avoided. However, compared to the IS method, the GLQ requires fewer samples~\cite{GLQ-1}. Thus, despite the large computational load of approximating $S(y)$ values~\cite{SGS-model-6}, the GLQ is still applicable to the estimation of $\mathrm{E}\left(y^{-1}\right)$.    
\end{rem}

\subsubsection{Gamma Series Based Estimator}
Both the IS and GLQ-based estimators suffer from low efficiency due to their requirements of numerous particles and the approximation of $S(y)$, respectively. In contrast, a more efficient estimator based on the GS is explained in detail in this section. We first show how the density $q^ \prime(y)$ can be represented by a series based on the inverse Gamma density (cf. Lemma~\ref{lem:GS}). Subsequently, a corresponding estimator of $\mathrm{E}\left(y^{-1}\right)$ is introduced as a result of Theorem~\ref{thm:series estimation}.
\begin{lem} \label{lem:GS}
For the SG$\alpha$S distribution with shape parameter $\alpha$, let $\alpha_1=\frac{\alpha}{2}$, then $q^ \prime(y)$ can be represented as an Inverse-Gamma-pdf-based series, i.e., 
\begin{equation} \label{eq:GS-1}
q^\prime(y)=\sum_{\xi=1}^{+\infty}c_\xi\frac{\Gamma(a_\xi)}{b^{a_\xi}}\mathrm{IG}(y;a_\xi,b)
\end{equation}
where 
\begin{align}
a_\xi&=\xi \alpha_1 +\frac{m}{2},\quad b=\frac{\eta}{2} \label{eq:GS-2}\\
c_\xi&=(-1)^{\xi+1}\frac{\Gamma(\xi\alpha_1+1)}{\pi \xi!}\mathrm{sin}(\xi\alpha_1\pi) \label{eq:GS-3}.
\end{align}
\end{lem}
\textit{Proof}: See Appendix~\ref{proof:lem:GS}.

\begin{thm} \label{thm:series estimation}
For the SG$\alpha$S distribution with shape parameter $\alpha$, $\mathrm{E}\left(y^{-1}\right)$ can be represented as the ratio of two Gamma function-based series sums as
\begin{equation} \label{eq:GS-4}
\mathrm{E}\left(y^{-1}\right)=\frac{\sum_{\xi=1}^{+\infty}c_\xi\frac{\Gamma(a_\xi+1)} {b^{(a_\xi+1)}}} {\sum_{\xi=1}^{+\infty}c_\xi\frac{\Gamma(a_\xi)} {b^{a_\xi}}} 
\end{equation}
 \emph{iff} both the series sums are convergent.
\end{thm}
\textit{Proof}: See Appendix~\ref{proof:thm:series estimation}.

\begin{rem}
The numerator and denominator series in equation~(\ref{eq:GS-4}) may not always converge. Hence, the estimator cannot be directly applied to the RKF-SG$\alpha$S, as it fails in cases when the estimator diverges.  
\end{rem}
 
\subsubsection{Hybrid Estimators}
Owing to the divergence of the GS-based estimator, two hybrid estimators, GSIS and GSGL, are further developed. When either/both the numerator or/and denominator series in~(\ref{eq:GS-4}) diverges, the GS estimator is replaced with the IS or GLQ-based methods. 

Considering the convergence ranges of the above-mentioned series are not available, convergence analysis requires testing. For this purpose, 
let
\begin{equation} \label{eq:Hybrid-1}
  r_\xi^{(1)}=c_\xi\frac{\Gamma(a_\xi+1)} {b^{(a_\xi+1)}}, \quad r_\xi^{(2)}=c_\xi\frac{\Gamma(a_\xi)} {b^{a_\xi}} 
\end{equation}
where $r_\xi^{(1)}$ and $r_\xi^{(2)}$ are the $\xi$-th elements of the numerator and denominator series, respectively.
Then, the convergence conditions of these two series can be written as~\cite{convergence-1}
\begin{equation} \label{eq:Hybrid-2}
 \sum_{\xi=\overline{\xi}-\tau_1}^{\overline{\xi}}\left|\frac{r_{\xi}^{(j)}}{\sum_{\xi^\prime=1}^{\xi}r_{\xi^\prime}^{(j)}}\right|<\varepsilon_1, \quad j=1,2
\end{equation}
where the threshold $\varepsilon_1$ is a small positive number. $\tau_1\geq1$ is a small positive integer used to ensure the series sums are stable when $\overline{\xi}-\tau_1 \leq \xi \leq \overline{\xi}$. Also, we set the largest value of $\overline{\xi}$ as $\Xi$. Following this, we can judge the convergence with the increase of $\overline{\xi}$. If $\overline{\xi}<\Xi$ and the condition in~(\ref{eq:Hybrid-2}) are satisfied, the estimation results are reliable. By contrast, we consider the estimator to diverge when $\overline{\xi}=\Xi$ and the convergence inequalities are violated. Then, either the IS or GLQ estimator is employed instead. The whole estimation method is shown in Algorithm~\ref{alg:GSIS and GSGL}.

\begin{algorithm} [h]
\caption{GSIS and GSGL estimation}\label{alg:GSIS and GSGL}
\textbf{Input:} $\eta, m, \alpha$

$b \gets \frac{\eta}{2}$;

\For {$\overline{\xi} \gets 1$ to $\Xi$}{
    
    Calculate $a_{\overline{\xi}}$ by~(\ref{eq:GS-2});
    
    Calculate $c_{\overline{\xi}}$ by~(\ref{eq:GS-3}); 
    
    Calculate $r_{\overline{\xi}}^{(1)},r_{\overline{\xi}}^{(2)}$ by~(\ref{eq:Hybrid-1});  

    \If{${\overline{\xi}}<\Xi$ and inequalities in~(\ref{eq:Hybrid-2}) are satisfied }{ 
       Estimate $E(y^{-1})$ by~(\ref{eq:GS-4});
       Break;
    }
    \If{${\overline{\xi}}=\Xi$}{
       Estimate $E(y^{-1})$ by~(\ref{eq:IS-1}) or~(\ref{eq:GL-1});
    }
 
}
\Return{$E(y^{-1})$}
\end{algorithm}

\subsection{Convergence Test for Fixed-point Iteration Method}\label{sec:convergence-test}

Under the RKF-SG$\alpha$S framework, the convergence of the fixed-point iteration to solve equation~(\ref{eq:VB}) needs to be investigated. In previous work~\cite{RSTKF-7}, $\mathbf{x}_{k|k}^{(t)}$ was solely tested. However, in this paper, we extend this approach by additionally detecting the convergence of the posterior covariance matrix $\mathbf{P}_{k|k}^{(t)}$ and the expectation of the scale function $\mathrm{E}^{(t)}\left(\kappa(\lambda_k)\right)$ to improve the detection reliability. Furthermore, we test the convergence of the latest $\tau_2>1$ items, as opposed to only the $t$-th item tested in~\cite{RSTKF-7}. The convergence conditions are expressed as
\begin{equation} \label{eq:fixed point convergence}
\sum_{t={\overline{t}}-\tau_2}^{\overline{t}} \frac{\mathbf{sum}\left(\mathbf{abs}\left(\boldsymbol{\Phi}^{(t)}-\boldsymbol{\Phi}^{(t-1)}\right)\right)}  {\mathbf{sum}\left(\mathbf{abs}\left(\boldsymbol{\Phi}^{(t)}\right)\right)} <\varepsilon_2
\end{equation}
where $\boldsymbol{\Phi} \in \{\mathbf{x}_{k|k}, \mathbf{P}_{k|k}, \mathrm{E}\left(\kappa(\lambda_k)\right) \}$ and $\varepsilon_2$ is a small positive number. In addition, the fixed-point iteration has a maximum number of iterations, denoted by $M$, after which the algorithm will terminate and output the current estimation results. The implementation pseudo-code for the proposed RKF-SG$\alpha$S is given in Algorithm~\ref{alg:RKF-SSG}.
\begin{algorithm} [h]
\caption{One Time Step of the Proposed RKF-SG$\alpha$S}\label{alg:RKF-SSG}
\textbf{Input:} $\mathbf{z}_k$, $\mathbf{\hat{x}}_{k-1|k-1}$, $\mathbf{P}_{k-1|k-1}$, $\mathbf{F}_k$, $\mathbf{H}_k$, $\mathbf{Q}_k$, $u_k$, $\mathbf{U}_k$, $\alpha$, $m$, $\Xi$, $\varepsilon_1$, $\varepsilon_2$, $\tau_1$, $\tau_2$, $M$, $N$, $L$

Calculate $\mathbf{P}_{k|k-1}$ using~(\ref{eq:Pk_k-1});

Initialisation: $\mathrm{E}^{(0)}\left(\mathbf{R}_k^{-1}\right)=\left(u_k-m-1\right)\left(\mathbf{U}_k\right)^{-1}$, $\mathrm{E}^{(0)}\left(\lambda_k^{-1}\right)=1$;

\For {$t \gets 0$ to $M-1$}{
    
    Calculate $\mathbf{\Tilde{R}}_k^{(t)}$ using~(\ref{eq:modified covariance});
    
    Calculate $\mathbf{\hat{x}}_{k|k}^{(t+1)}$ and $\mathbf{P}_{k|k}^{(t+1)}$ using~(\ref{eq:KF update}); 

    Calculate $\mathbf{B}_k^{(t+1)}$ and $\eta$ using~(\ref{eq:eta and B});

    Calculate $\mathrm{E}^{(t+1)}\left(\lambda_k^{-1}\right)$ using one of the scale function estimators in section~\ref{sec:scale-estimator};

    Calculate $\mathbf{b}_k^{(t+1)}$, $\mathbf{D}_k^{(t+1)}$, $u_k^{(t+1)}$ 
     and $\mathbf{U}_k^{(t+1)}$ using~(\ref{eq:parameters of IW});

    Calculate $\mathrm{E}^{(t+1)}\left(\mathbf{R}_k^{-1}\right)$ using~(\ref{E_Rk}); 
    
    \If{The convergence condition in~(\ref{eq:fixed point convergence}) is satisfied}{
       Terminate the iteration;
    }
}
\Return{$\mathbf{\hat{x}}_{k|k}=\mathbf{\hat{x}}_{k|k}^{(M)}$, $\mathbf{P}_{k|k}=\mathbf{P}_{k|k}^{(M)}$}
\end{algorithm}

\section{NUMERICAL SIMULATIONS} \label{sec: experiments}
\subsection {Target Tracking Models and Noises} \label{exp:model}
In this section, we introduce the target tracking model used in the experimental analysis. The target moves uniformly in a straight line, and the state-space model is 
given by~(\ref{eq:statespace_1})
\begin{align}
\mathbf{F}_k=
\begin{bmatrix}
\mathbf{I}_2 & \triangle t \mathbf{I}_2\\
\mathbf{0} & \mathbf{I}_2
\end{bmatrix}, 
&\quad 
\mathbf{H}_k=
\begin{bmatrix}
\mathbf{I}_2 & \mathbf{0}
\end{bmatrix}
\end{align}
with the observation interval of $\triangle t=1$. Also, {the initial target state is chosen from $\mathcal{N}(\mathbf{x}_{0|0}, \mathbf{P}_{0|0})$  randomly, where $\mathbf{x}_{0|0}=\left[0,0,10,10\right]^T$, $\mathbf{P}_{0|0}=\mathrm{diag}\left(\left[25, 25, 2, 2\right]\right)$.} Besides, the process noise $\mathbf{w}_k$ follows the Gaussian distribution, of which the covariance matrix is given by
\begin{equation} \label{eq:nominal Q} 
 \mathbf{Q}_k= 0.1*\overline{\mathbf{Q}}, 
\quad
\overline{\mathbf{Q}}=  
\begin{bmatrix}
\frac{\triangle t^3}{3}\mathbf{I}_2 & \frac{\triangle t^2}{2}\mathbf{I}_2 \\
\frac{\triangle t^2}{2}\mathbf{I}_2 & \triangle t\mathbf{I}_2
\end{bmatrix} 
.
\end{equation}
Then, three kinds of heavy-tailed measurement noises are selected---the Gaussian Mixture (GM), ST and SG$\alpha$S noises. Among them, the GM noise has been employed to test the filtering performance of RKFs in many references~\cite{RSTKF-1,RSTKF-3,RKF-GSM-1}. In contrast, we also consider the ST and SG$\alpha$S noises thanks to their capability to fit the practical noises. The GM noise~\cite{RKF-GSM-1,RKF-GSM-2} is written as
\begin{equation}
\mathbf{GM} = 
\begin{cases}
 \mathcal{N}(\mathbf{0},\mathbf{\overline{R}}) \quad {\rm with \ probability} \quad 0.9\\
 \mathcal{N}(\mathbf{0}, U \mathbf{\overline{R}}) \quad {\rm with \ probability} \quad 0.1,
\end{cases}
\end{equation}
where $\mathbf{\overline{R}}=10\mathbf{I}_2$ is the nominal covariance matrix. Also, $U$ is the augment factor which is taken from a vector $[5,$ $ 10,$ $ 10^2,$ $ 10^3,$ $ 10^4,$ $ 10^5,$ $ 10^6,$ $ 10^7,$ $ 10^8]$. Contrarily, the SG$\alpha$S and ST noises have the same scale matrix $\mathbf{\overline{R}}$, and their shape parameters, $\alpha$ and $v$, take values from [0.3, 0.5, 0.7, 0.9, 1.1, 1.3, 1.5, 1.7, 1.85] and [0.3, 0.5, 0.7, 0.9, 1.2, 1.7, 2.5, 3.5, 6], respectively. The whole tracking process lasts for {300} seconds, and 100 Monte Carlo runs are simulated.

\subsection {Benchmark Filters} \label{Bench filters}
The RKF-SG$\alpha$S is compared with 3 kinds of heavy-tailed-distribution-based RKFs in the following simulations. The first two filters are the RKF based on the variance Gamma distribution (RKF-VG)~\cite{RKF-GSM-1} and the RSTKF~\cite{RSTKF-6}. Lastly, our improved version of RKF-SL is also used as a comparison method. Besides, the standard KF with true noise covariance matrices (KFTNCM)~\cite{RKF-GSM-2} is employed for reference.

For the RKF-SG$\alpha$S, the performance of the four variants is evaluated, including RKF-SG$\alpha$S-IS, RKF-SG$\alpha$S-GLQ, RKF-SG$\alpha$S-GSIS, RKF-SG$\alpha$S-GSGL. Then, for the GS estimator, $\Xi=30$, $\varepsilon_1=10^{-2}$ and $\tau_1=4$. Moreover, we set $M=50$, $\varepsilon_2=10^{-2}$ and $\tau_2=4$  for all the RKF-based filters.

\subsection {Estimation of RKF Parameters} \label{sec:noise modelling}
In this section, we employ the EM or MLE method to initialise the shape parameters of the GSM distributions, $u_k$, and $\mathbf{U}_k$. Compared with manually setting these RKF parameters as in the previous work~\cite{RKF-GSM-1}, this can alleviate the influence of the model errors and provide a fairer comparison among these RKFs. {
Considering the two-dimensional isotropic measurement noise, we simply assume that the covariance matrix of the GSM distributions is a scalar matrix. Following this, the parameter estimation of the two-dimensional GSM distributions can be addressed as in the univariate case. Specifically, we can first estimate the parameters of a univariate GSM distribution based on the one-dimensional squeezed measurement noise samples, and then the estimated shape and variance parameters equal the shape parameter and the diagonal elements of the covariance matrix of the two-dimensional GSM distribution, respectively.} 

For the SG$\alpha$S distribution, the EM algorithm is employed~\cite{SGS-4}. Specifically, the EM iteration number is $200$, and the estimates of the last $15$ iterations are averaged for the final estimate. Besides, to sample the posterior distribution of the Weibull random variable under this EM parameter estimation framework, a rejection sampling method is used and the sample number is $2000$. In contrast to the case of SG$\alpha$S mentioned above, the MLE algorithm is utilised for the parameter estimation of the VG, SL and ST distributions. As we only consider the zero-mean symmetric noise, the values of both the mean and skewness vectors for all the distributions are assumed to be known. The measurement noise sample size is 1000, and 1000 Monte Carlo runs are taken. The shape parameter ($\alpha$ or $v$) is estimated by averaging the results. Lastly, 1000 scale matrix estimation results are modelled by the IW distribution, and the MLE method is used to estimate its parameters $u_k$ and $\mathbf{U}_k$.

\subsection {Performance Evaluation of RKF-SG$\alpha$S Variants} \label{sec: experiment-property}
\begin{figure}[h!]
\centering
\includegraphics[width=\linewidth]{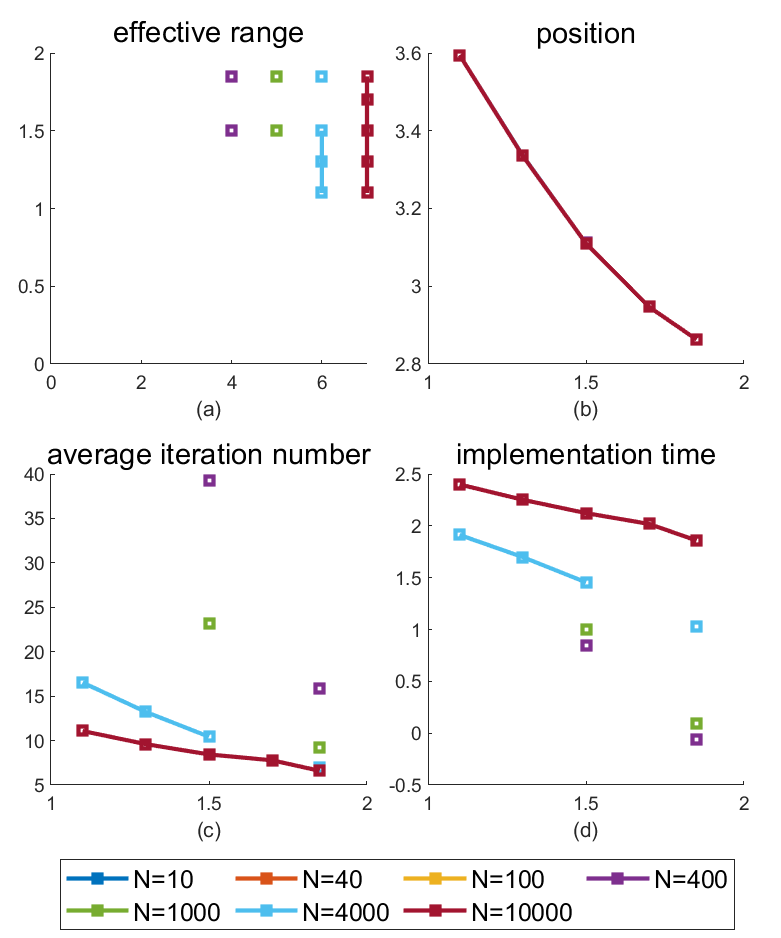}
\caption{The target tracking results of the RKF-SG$\alpha$S-IS under different particle numbers $N$. For (a), the x-axis is the particle number index and the y-axis represents the shape parameter value. In comparison, for (b)-(d), the x-axis is the shape parameter value, and the y-axis represents the estimation RMSE, iteration number and log(time), respectively. (a) depicts the shape parameter ranges, where the filter can track the target steadily. By contrast, (b) show the position estimation RMSEs. Also, (c) describes the average fixed-point iteration numbers. Besides, the average implementation time of one Monte Carlo run is shown in (d).}
\label{fig:IS}
\end{figure}

\begin{figure}[h!]
\centering
\includegraphics[width=\linewidth]{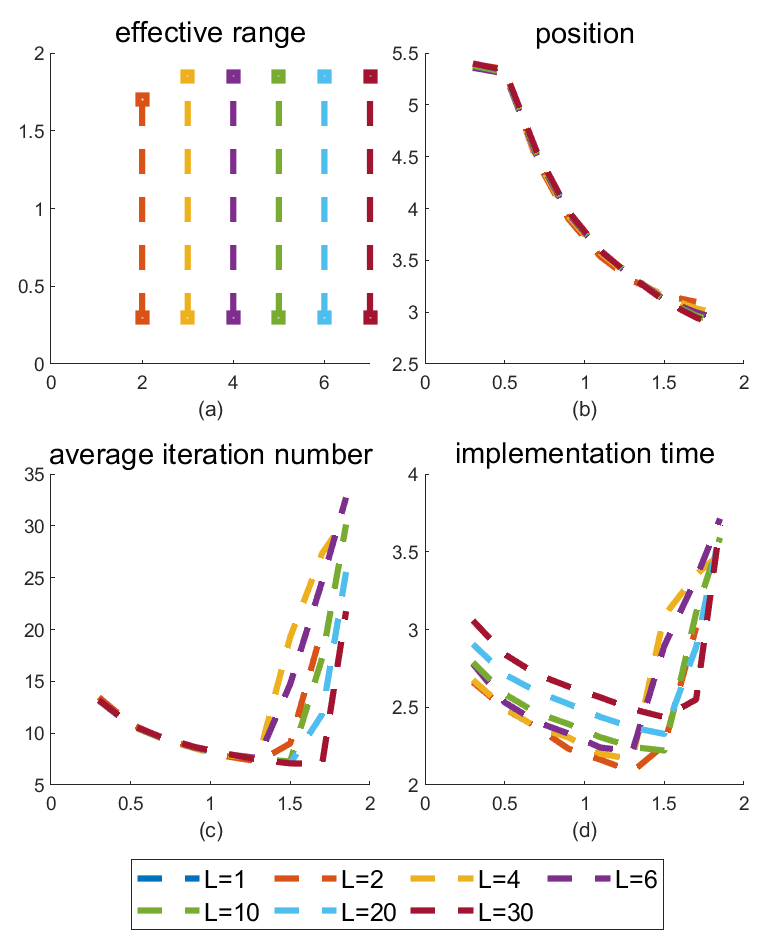}
\caption{The target tracking results of the RKF-SG$\alpha$S-GLQ under different root numbers $L$. For (a), the x-axis is the root number index and the y-axis represents the shape parameter value. In comparison, for (b)-(d), the x-axis is the shape parameter value, and the y-axis represents the estimation RMSE, iteration number and log(time), respectively. (a) depicts the shape parameter ranges, where the filter can track the target steadily. By contrast, (b) show the position estimation RMSEs. Also, (c) describes the average fixed-point iteration numbers. Besides, the average implementation time of one Monte Carlo run is shown in (d).}
\label{fig:GL}
\end{figure}

\begin{figure}[h!]
\centering
\includegraphics[width=\linewidth]{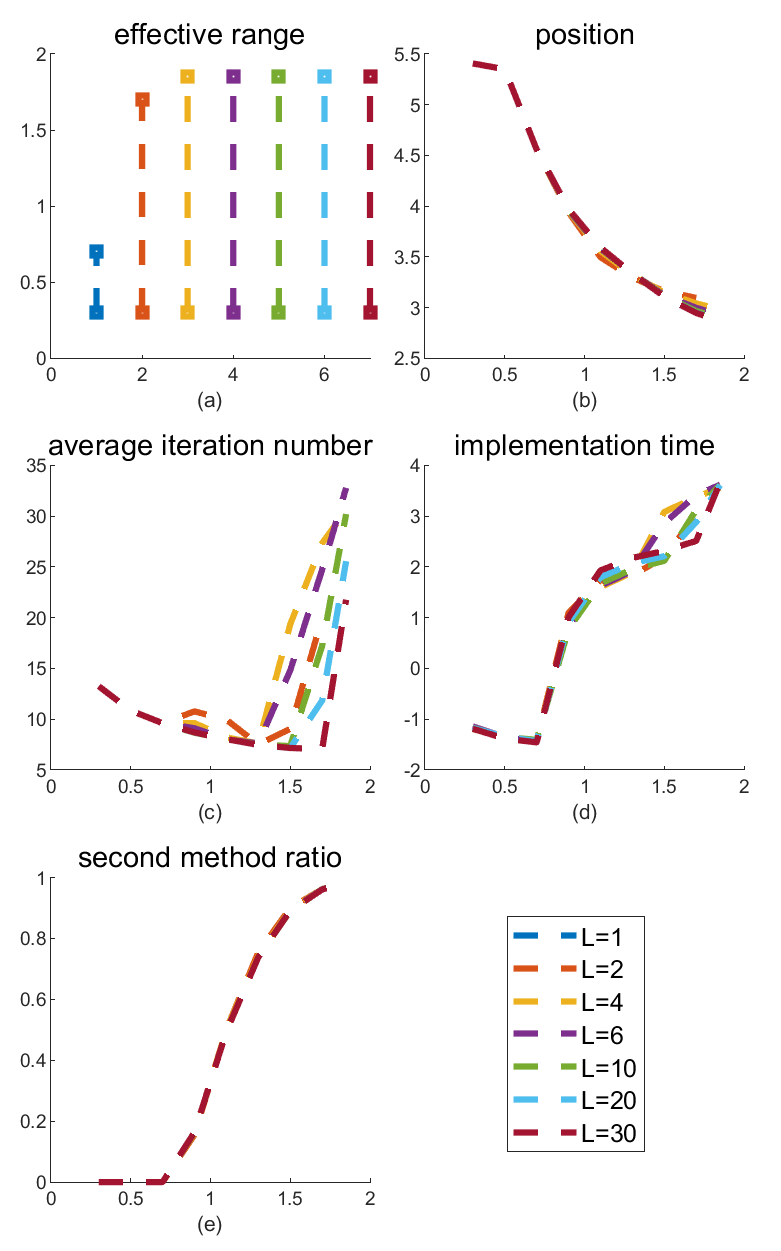}
\caption{The target tracking results of the RKF-SG$\alpha$S-GSGL under different root numbers $L$. For (a), the x-axis is the root number index and the y-axis represents the shape parameter value. In comparison, for (b)-(e), the x-axis is the shape parameter value, and the y-axis represents the estimation RMSE, iteration number, log(time) and ratio value, respectively. (a) depicts the shape parameter ranges, where the filter can track the target. By contrast, (b) shows the position estimation RMSEs. Also, (c) describes the average fixed-point iteration numbers. Then, the average implementation time of one Monte Carlo run is shown in (e). Besides, (f) paints the ratio of the iterations where the GS method is replaced with the GLQ estimator.}
\label{fig:GSGL}
\end{figure}

\begin{figure}[h!]
\centering
\includegraphics[width=\linewidth]{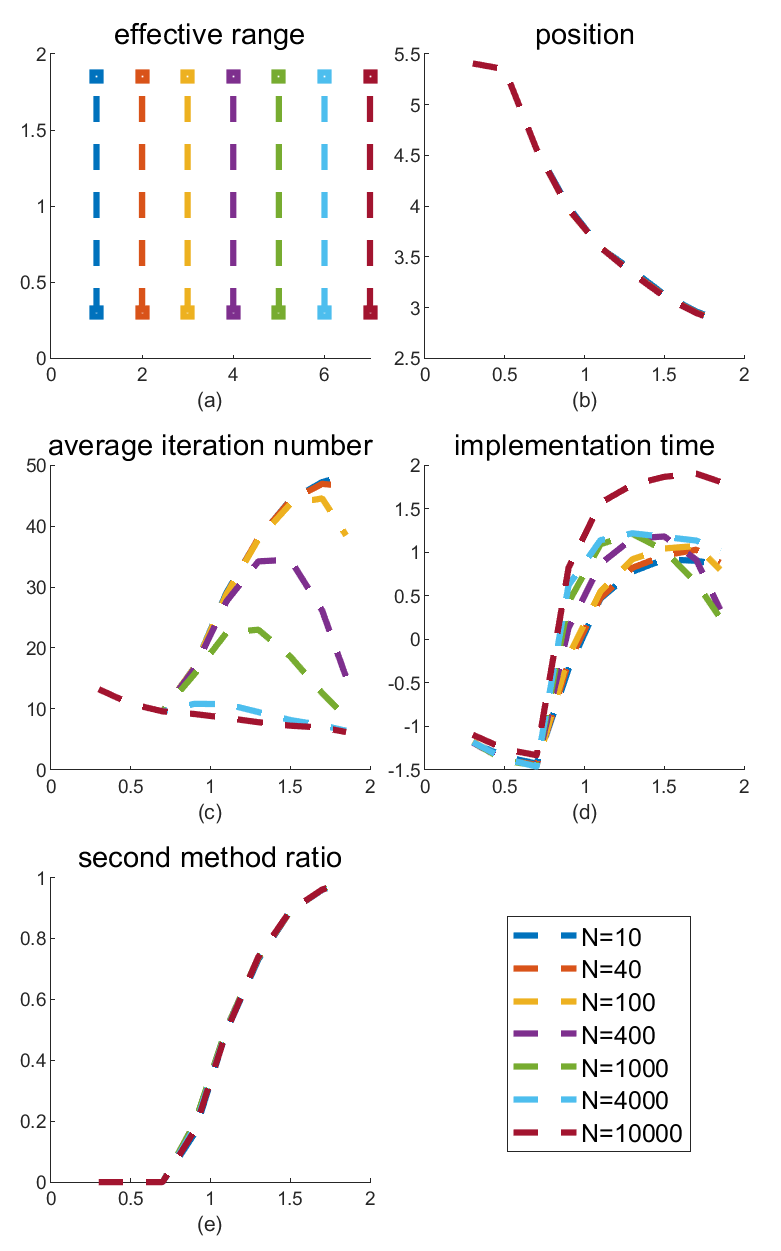}
\caption{The target tracking results of the RKF-SG$\alpha$S-GSIS under different particle numbers $N$. For (a), the x-axis is the root number index and the y-axis represents the shape parameter value. In comparison, for (b)-(e), the x-axis is the shape parameter value, and the y-axis represents the estimation RMSE, iteration number, log(time) and ratio value, respectively. (a) depicts the shape parameter ranges, where the filter can track the target. By contrast, (b) shows the position estimation RMSEs. Also, (c) describes the average fixed-point iteration numbers. Then, the average implementation time of one Monte Carlo run is shown in (d). Besides, (e) depicts the ratio of the iterations where the GS method is replaced with the IS estimator.}
\label{fig:GSIS}
\end{figure}

\begin{figure}[h!]
\centering
\includegraphics[width=\linewidth]{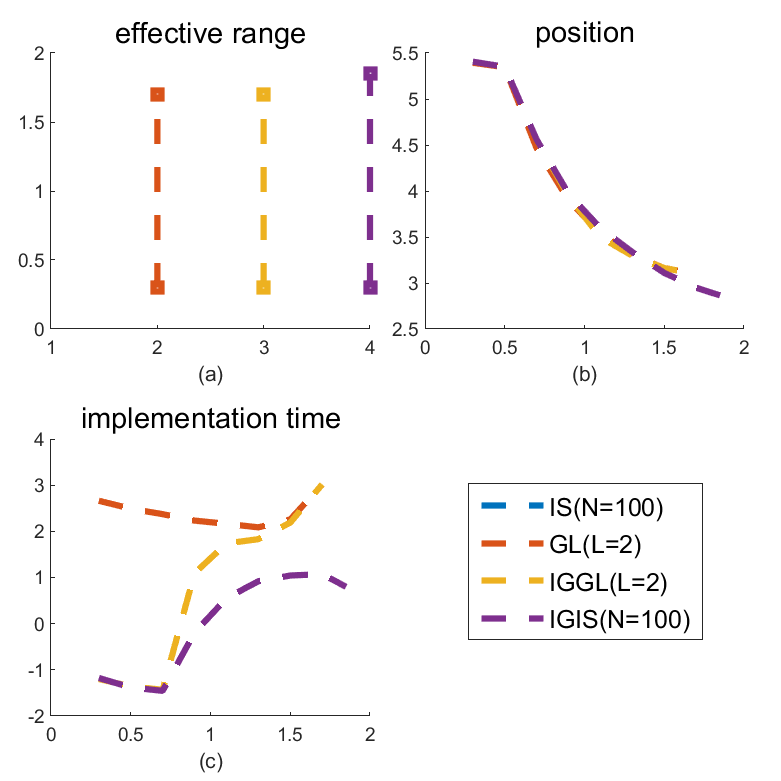}
\caption{{The target tracking results of the four RKF-SG$\alpha$S variants. For (a), the x-axis is the variant index and the y-axis represents the shape parameter value. In comparison, for (b)-(c), the x-axis is the shape parameter value, and the y-axis represents the estimation RMSE and log(time), respectively. (a) depicts the shape parameter ranges, where the filter can track the target. By contrast, (b) shows the position estimation RMSEs, and the average implementation time of one Monte Carlo run is shown in (c).}}
\label{fig:estimators}
\end{figure}

Based on the parameter estimation results, the properties of our proposed RKF-SG$\alpha$S variants are investigated in this section. The particle number $N$ and the root number $L$ are taken from the vectors [10, 40, 100, 400, 1000, 4000, 10000] and [1, 2, 4, 6, 10, 20, 30], respectively. Also, the measurement noises come from SG$\alpha$S distributions, and the parameter selection is explained in section~\ref{exp:model}. 

Figure~\ref{fig:IS} shows the experimental results of RKF-SG$\alpha$S-IS. (1) From Figure~\ref{fig:IS}-(a), larger particle numbers allow for the broader effective ranges of $\alpha$ { and more stable filtering results}. The heavier-tailed distributions contain larger probability spaces, and then more samples are needed for the IS approximation. (2) The position estimation RMSEs in Figure~\ref{fig:IS}-(b) are not influenced by $N$ within the effective ranges of $\alpha$. This suggests that the larger estimation errors of $\mathrm{E}\left(\kappa(\lambda_k)\right)$ caused by small $N$ have little impact on the estimation of the position marginal distribution. (3) From Figure~\ref{fig:IS}-(c), the average iteration number is inversely proportional to $N$, as the imprecise estimate of $\mathrm{E}\left(\kappa(\lambda_k)\right)$ causes the EM estimator to fluctuate around the local optimum and cannot satisfy the convergence inequalities in~(\ref{eq:fixed point convergence}). Thus, the convergence process is delayed. (4) From Figure~\ref{fig:IS}-(d), the execution time is proportional to the sample size. Although large $N$ can reduce the iteration number, the filter with small $N$ is still more efficient because of the shorter implementation time on the IS. 

The tracking performance of RKF-SG$\alpha$S-GLQ is plotted in Figure~\ref{fig:GL}. (1) The effective $\alpha$ range shown in Figure~\ref{fig:GL}-(a) is related to the root number. For $L=1$, the filter fails on the whole $\alpha$ range, because the GLQ cannot estimate the integral. By contrast, for $L=2$, the effective range is large. However, when $\alpha$ is large, the target is lost. For the light-tailed case, the probability space of $S(y)$ concentrates on a small range, and then the GLQ produces rough estimates. In comparison, the filter is stable on the whole tested ranges when $L\geq 4$. (2) The position estimation (Figure~\ref{fig:GL}-(b)) precision cannot be improved by larger $L$ although the GLQ can produce more precise estimates of $\mathrm{E}\left(\kappa(\lambda_k)\right)$. (3) From Figure~\ref{fig:GL}-(c), the convergence of the fixed-point iteration method is influenced by both $\alpha$ and $L$. When $\alpha$ is large, smaller $L$ requires more iterations due to the fluctuation around the local optimum. By contrast, when $\alpha<1.3$, different values of $L$ need similar iteration numbers since the GLQ estimation with small $L$ is still precise in the heavy-tailed cases. (4) From Figure~\ref{fig:GL}-(d), the relation between the execution time and $L$ is complicated. When $\alpha$ is small, the filter with small $L$ is more efficient. Nevertheless, for large $\alpha$, its efficiency degrades caused by the increased number of iterations.

The simulation results of the RKF-SG$\alpha$S-GSGL are plotted in Figure~\ref{fig:GSGL}. (1) For this hybrid-estimator-based filter, the ratio of the second method, GLQ, is added to  Figure~\ref{fig:GSGL}-(e). For small $\alpha$, the ratio is $0$. However, with the increase of $\alpha$, the ratio of the GLQ gradually rises towards 1. This suggests that the GS estimator is more stable when the noise is heavy-tailed. (2) The effective range is related to $L$. For $L=1$, although the RKF-SG$\alpha$S-GLQ always fails in Figure~\ref{fig:GL}-(a), the effective range of this hybrid filter is $\alpha \leq 0.7$, where the GS estimator always converges. By contrast, for $L>1$, the effective range is the same as that of the RKF-SG$\alpha$S-GLQ. (3) The position estimation results are always precise within the effective ranges. (4) The iteration number and the execution-time lines overlap for small $\alpha$, where the hybrid filter mainly relies on the GS estimator. However, from Figure~\ref{fig:GL}-(d) and~\ref{fig:GSGL}-(d), the RKF-SG$\alpha$S-GSGL is more efficient than the RKF-SG$\alpha$S-GLQ in heavy-tailed cases, which is caused by the smaller computation loads of the GS estimator. By contrast, with the increase of $\alpha$, their results gradually become similar due to the rising ratio of the GLQ estimator.

The performance of RKF-SG$\alpha$S-GSIS is evaluated in Figure~\ref{fig:GSIS}. (1) The ratio of the second method, the IS method, is similar to that of the RKF-SG$\alpha$S-GSGL in Figure~\ref{fig:GSGL}-(e). (2) The hybrid filter is stable on the whole $\alpha$ range, even when $N$ is small. The GS estimator is unstable for large $\alpha$, whilst the IS method with few particles can produce precise results in this case (cf. Figure~\ref{fig:IS}). So, the complementary advantages of the GS and IS strategies extend the effective range of the filter. (3) The hybrid filter can work in the $\alpha$ ranges where both the IS and GS-based filters fail. For example,  from Figures~\ref{fig:IS}-(a) and~\ref{fig:GSGL}-(a), the IS estimator with $N=100$ {fails in the whole tested range, and GS estimator is effective for only $\alpha \leq 0.7$.} However, Figure~\ref{fig:GSIS}-(a) shows the hybrid filter with $N=100$ can also work when {$\alpha >0.7$}. This suggests that although the GS estimator cannot converge at every iteration {when $\alpha > 0.7$}, the IS method is effective when the GS diverges. (4) All the position estimation RMSEs are close. (5) Based on Figures~\ref{fig:IS}-(c-d) and~\ref{fig:GSIS}-(c-d), the hybrid filter is more efficient than the RKF-SG$\alpha$S-IS when $\alpha$ is small because of the high ratio of the GS estimator as shown in Figure~\ref{fig:GSIS}-(e). By contrast, their performances are similar when $\alpha$ is large.

{Figure~\ref{fig:estimators} provides a filtering performance comparison among these four variants of RKF-SG$\alpha$S, where $N=100, L=2$. (1) From Figure~\ref{fig:estimators}-(a), the IGIS-based filter presents the broadest effective filtering range owing to the complementary advantages of the IS and IG estimators. (2) Within the effective scales, all the filters produce similar filtering results in Figure~\ref{fig:estimators}-(b). (3) Figure~\ref{fig:estimators}-(c) shows the RKF-SG$\alpha$S-IGIS requires the smallest computational loads over the whole tested scale though its implementation time is similar to that of the IGGL-based filter for small $\alpha$.}

In summary, for the effective range, the performance of RKF-SG$\alpha$S-IS is worse than the other filters, and numerous particles are needed for small $\alpha$. Conversely, the position estimation results of all four proposed strategies are similar in the effective $\alpha$ ranges. However, benefiting from the complementary advantages of the GS and IS estimators, the RKF-SG$\alpha$S-GSIS with few particles performs steadily and then becomes the most efficient.

\subsection {RKF-SG$\alpha$S vs. Benchmark Filters} \label{sec: experiment-comparison}
\begin{figure}[h]
\centering
\includegraphics[width=\linewidth]{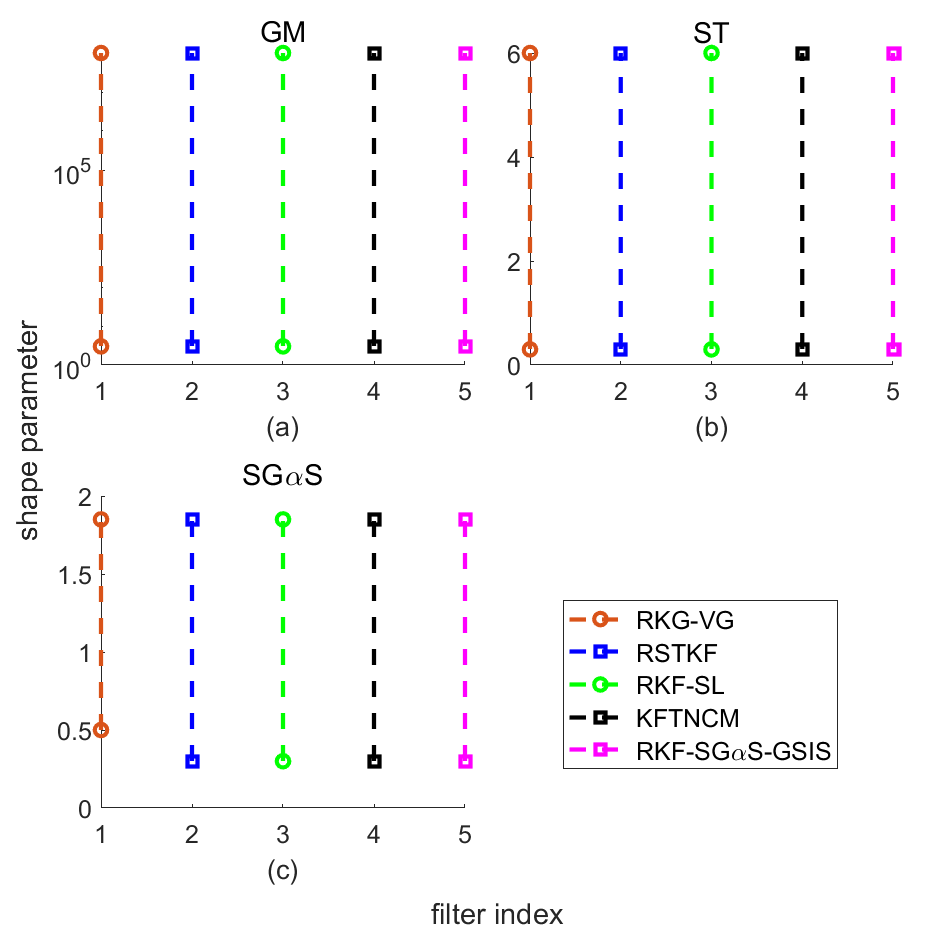}
\caption{The effective shape parameter ranges of different filters under the three kinds of measurement noises. The titles of the subplots are the noise distributions, {the x-axis is the filter index, and the y-axis represents the shape parameter value.} Also, for simplicity, the parameter $U$ of GM distribution is defined as the shape parameter.}
\label{fig:effective range}
\end{figure}

\begin{figure}[h]
\centering
\includegraphics[width=\linewidth]{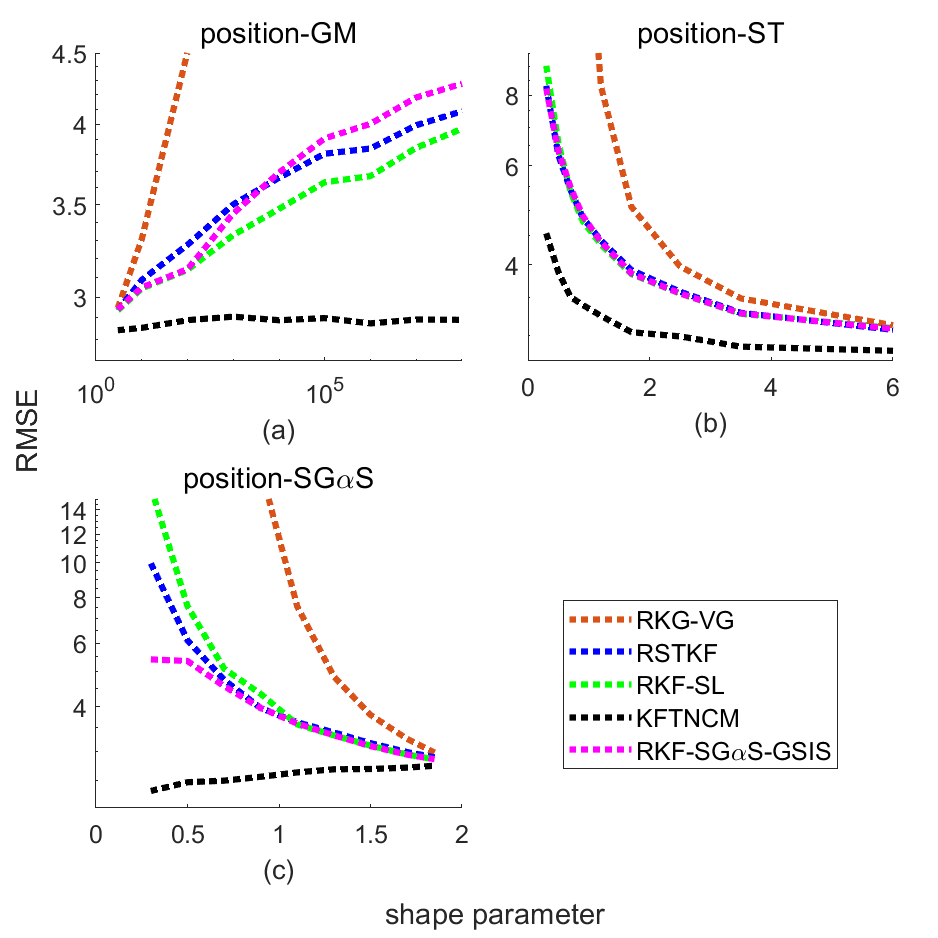}
\caption{The position estimation RMSEs of different filters under the three kinds of measurement noises. The titles of the subplots follow the format: position-noise distributions, {the x-axis is the shape parameter value, and the y-axis represents position estimation RMSEs.} Also, for simplicity, the $U$ of GM distribution is defined as the shape parameter.}
\label{fig:position}
\end{figure}

\begin{figure}[h]
\centering
\includegraphics[width=\linewidth]{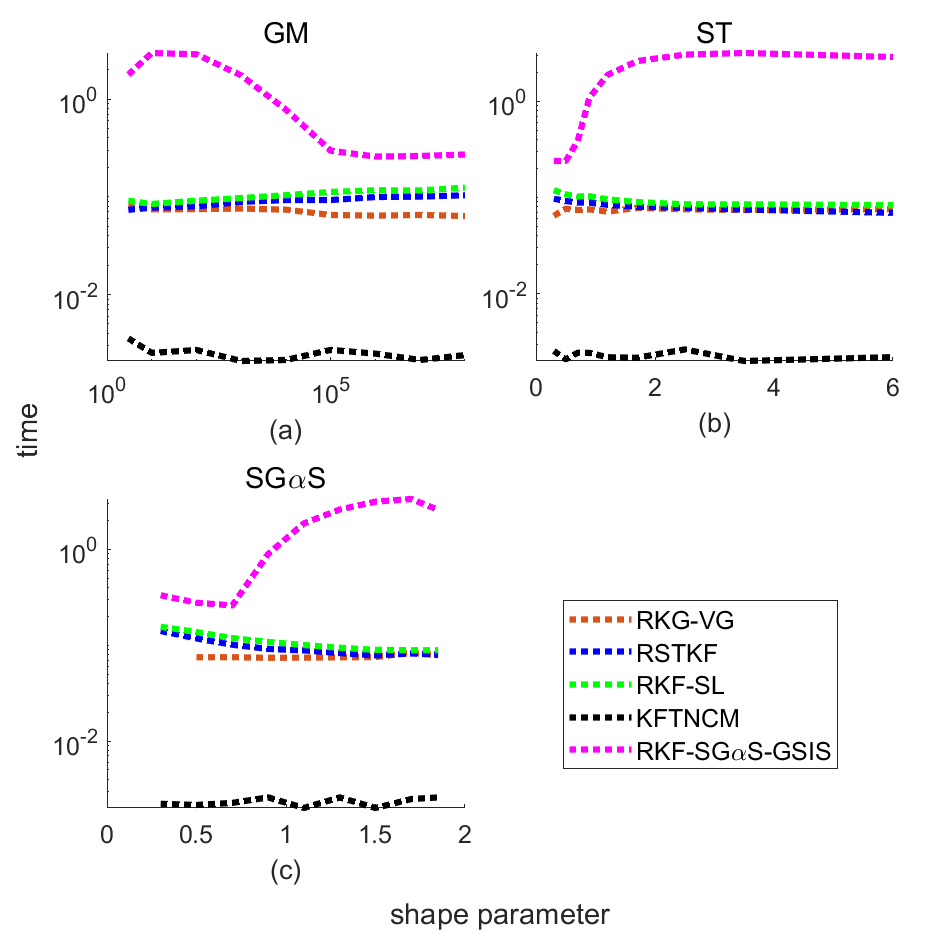}
\caption{The average execution time of one Monte Carlo run of different filters under the three kinds of measurement noises. The titles of the subplots are the noise distributions, {the x-axis is the shape parameter value, and the y-axis represents the implementation time.} Also, for simplicity, the $U$ of GM distribution is defined as the shape parameter.}
\label{fig:computational time}
\end{figure}

\begin{figure}[h!]
\centering
\includegraphics[width=\linewidth]{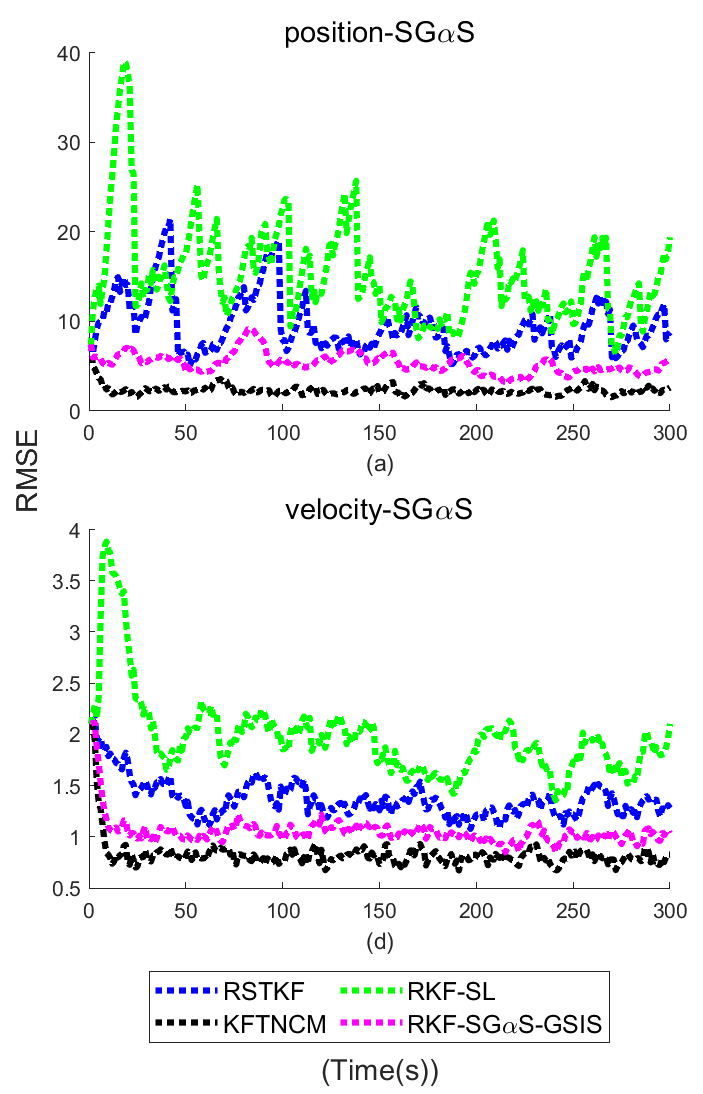}
\caption{{The filtering error dynamics over time. For (a) and (b), the x-axis is the tracking times, and the y-axis represents the position/velocity estimation RMSEs.}}
\label{fig:TRMSE}
\end{figure}

Following the thorough analysis of the RKF-SG$\alpha$S in the previous sections, {the comparison between the RKF-SG$\alpha$S-GSIS and the benchmark filters}, listed in Section~\ref{Bench filters}, is presented in this section. For efficiency, we set {$N=100$}. Also, three kinds of measurement noises are employed, and their parameter selection is made as explained in section~\ref{exp:model}. 

In Figure~\ref{fig:effective range}, the effective range results are shown. The RKF-VG fails when the shape value is small under the SG$\alpha$S noises. On the contrary, all other filters behave stably over the whole tested range.

The position estimation results are depicted in Figure~\ref{fig:position}. Under light-tailed noise, all the filters produce similar results. However, under the heavy-tailed noises, their performances vary. Compared to RKF-VG, the RKF-SG$\alpha$S always performs better under different measurement noises. Also, the RKF-SL obtains a more precise estimation than RKF-SG$\alpha$S under the GM noise, but worse results under SG$\alpha$S noises. Further, the RSTKF and RKF-SG$\alpha$S achieve comparable estimates under GM and ST noises. Nonetheless, the RKF-SG$\alpha$S becomes superior under the heavy-tailed SG$\alpha$S noise.

{Figure~\ref{fig:computational time} plots the execution time of different filters. Under light-tailed noise, the computational cost of the RKF-SG$\alpha$S is higher. Nevertheless, their performance becomes similar in the heavy-tailed cases because the high ratio of the GS estimator increases the filtering efficiency.}

{Figure~\ref{fig:TRMSE} presents the filtering error dynamics of the position and velocity estimation, for SG$\alpha$S noise with characteristic exponent $\alpha=0.3$. The RKF-VG fails in this scenario, so no corresponding filtering performance is plotted. Also, all the other filters produce stable estimation results over time, and the RKF-SG$\alpha$S achieves the best results.}

\section{Conclusion} \label{sec:conclusion}
In this paper, {considering linear models with the SG$\alpha$S noise}, a robust Kalman filter framework based on the SG$\alpha$S distribution is proposed. 
First, we present four RKF-SG$\alpha$S variants based on the different MMSE estimators of the scale function. Then their properties are investigated. The experimental results indicate that the RKF-SG$\alpha$S-GSIS obtains the most efficient performance thanks to the complementary advantages of the GS and IS estimators.
Second, the RKF-SG$\alpha$S is compared with {several heavy-tailed-distribution-based} RKFs. From the simulations, the RKF-SG$\alpha$S produces similar estimates under the GM and ST noises but obtains the best performance under the heavy-tailed SG$\alpha$S noise. Considering the importance of the SG$\alpha$S noise, this superiority over benchmark filters is significant. Besides, although our RKF-SG$\alpha$S has higher computational complexity than the benchmark filters, their execution times become closer in particularly heavy-tailed cases.

In the future, we will consider applying the proposed framework to the state-space models with both heavy-tailed signal and measurement noises. In this work, only the measurement noise is considered heavy-tailed since the parameter estimation of the one-step prediction noise is more difficult compared to that of the measurement noise. However, it is straightforward to apply the RKF-SG$\alpha$S to dynamic models with heavy-tailed state noise.

\appendix
\subsection{Proof of Theorem~\ref{thm:IS}} \label{proof:thm:IS}
\begin{prf}
According to~(\ref{eq:Ey^{-1}}), to calculate $\mathrm{E}\left(y^{-1}\right)$, two integrals, $\int_0^{+\infty}y^{-1}q^ \prime(y)dy$ and $\int_0^{+\infty}q^ \prime(y)dy$, need to be approximated. 

Based on the IS method and~(\ref{eq:q_y}), these two integrals can be approximated based on the samples $y_i$ from $S(y)$. We rewrite
\begin{align*}
\int_0^{+\infty}y^{-1}q^ \prime(y)dy &\approx \frac{1}{N} \sum_{i=1}^N y_i^{-\frac{m}{2}-1}\mathrm{exp}\left(-\frac{\eta}{2y_i}\right) \\ 
\int_0^{+\infty}q^ \prime(y)dy &\approx \frac{1}{N} \sum_{i=1}^N y_i^{-\frac{m}{2}}\mathrm{exp}\left(-\frac{\eta}{2y_i}\right). 
\end{align*}
Then $\mathrm{E}\left(y^{-1}\right)$ in~(\ref{eq:Ey^{-1}}) can be obtained by using the expressions above and 
Theorem~\ref{thm:IS} holds \qedsymbol.
\end{prf}


\subsection{Proof of Theorem~\ref{thm:GL}} \label{proof:thm:GL}
\begin{prf}

In order to approximate the integrals in~(\ref{eq:Ey^{-1}}), first let $y=\frac{\eta}{2x}$, then
\begin{align*}
\begin{split}
\int_0^{+\infty}q^ \prime(y)dy&=\left(\frac{\eta}{2}\right)^{-\frac{m}{2}+1} \\
 &\times\int_0^{+\infty}\mathrm{\mathrm{exp}}(-x)x^{\frac{m}{2}-2}S\left(\frac{\eta}{2x}\right)dx
\end{split} 
\\
\begin{split}
\int_0^{+\infty}y^{-1}q^ \prime(y)dy&=\left(\frac{\eta}{2}\right)^{-\frac{m}{2}} \\
&\times\int_0^{+\infty}\mathrm{\mathrm{exp}}(-x)x^{\frac{m}{2}-1}S\left(\frac{\eta}{2x}\right)dx
\end{split} 
\end{align*}
Recalling~(\ref{eq:Ey^{-1}}) and replacing the expressions above in~(\ref{eq:Ey^{-1}}), we get
\begin{equation*}
\mathrm{E}\left(y^{-1}\right)=\frac{\int_0^{+\infty}x\mathrm{\mathrm{exp}}(-x)f(x)dx}{\frac{\eta}{2}\int_0^{+\infty}\mathrm{\mathrm{exp}}(-x)f(x)dx}
\end{equation*}
where $f(x)=x^{\frac{m}{2}-2}S(\frac{\eta}{2x})$. Then the GLQ method in Theorem~\ref{thm:GL} holds \qedsymbol.
\end{prf}

\subsection{Proof of Lemma~\ref{lem:GS}} \label{proof:lem:GS}
\begin{prf}
For the SG$\alpha$S with shape parameter $\alpha$, let $\alpha_1=\frac{\alpha}{2}$, then $S(y)$ can be represented as a series~\cite{SGS-5}
\begin{equation*}
S(y)=-\frac{1}{\pi y}\sum_{\xi=1}^{+\infty} \frac{\Gamma(\xi\alpha_1+1)}{\xi!} \left(-y^{-\alpha_1}\right)^\xi \mathrm{sin}{(\xi\alpha_1\pi)} 
\end{equation*}
Then, replacing $S(y)$ with the expression above in~(\ref{eq:q_y}),
\begin{equation} \label{q_y series}
 q^ \prime (y)=\sum_{\xi=1}^{+\infty}c_\xi y^{-a_\xi-1}\mathrm{\mathrm{exp}}\left(-\frac{b}{y}\right)
\end{equation}
where $a_\xi$, $b$ and $c_\xi$ are defined in~(\ref{eq:GS-2}) and~(\ref{eq:GS-3}). As $y^{-a_\xi-1}\mathrm{\mathrm{exp}}\left(-\frac{b}{y}\right)$ is proportional to an IG distribution, equation~(\ref{q_y series}) can be reformulated as in~(\ref{eq:GS-1}) which finalises the proof of Lemma~\ref{lem:GS} \qedsymbol.
\end{prf}

\subsection{Proof of Theorem~\ref{thm:series estimation}} \label{proof:thm:series estimation}
\begin{prf} 

Based on equation~(\ref{eq:GS-1}), we can represent the two integrals in~(\ref{eq:Ey^{-1}}) with two series respectively as
\begin{equation} \label{eq:qy-series-1}
\int_0^{+\infty}q^ \prime(y)dy = \sum_{\xi=1}^{+\infty}c_\xi\frac{\Gamma(a_\xi)} {b^{a_\xi}}\int_0^{+\infty}\mathrm{IG}(y;a_\xi,b)dy    
\end{equation}
\begin{equation} \label{eq: qy-series-2}
 \int_0^{+\infty}y^{-1}q^ \prime(y)dy = \sum_{\xi=1}^{+\infty}c_\xi\frac{\Gamma(a_\xi)} {b^{a_\xi}}\int_0^{+\infty}\frac{\mathrm{IG}(y;a_\xi,b)}{y}dy   
\end{equation}
assuming the above series are convergent. As $\int_0^{+\infty}\mathrm{IG}(y;a_\xi,b)dy=1$, from~(\ref{eq:qy-series-1}), we have
\begin{equation} \label{eq: qy-series-3}
\int_0^{+\infty}q^ \prime(y)dy = \sum_{\xi=1}^{+\infty}c_\xi\frac{\Gamma(a_\xi)} {b^{a_\xi}}.
\end{equation}
Following this, since $y\sim\mathrm{IG}(y;a_\xi,b)$, $y^{-1}\sim\mathrm{G}(y;a_\xi,b)$, then
\begin{equation} \label{eq: qy-series-4}
\int_0^{+\infty}\frac{\mathrm{IG}(y;a_\xi,b)}{y}dy= \int_0^{+\infty}y\mathrm{G}(y;a_\xi,b)dy=\frac{a_\xi}{b}.
\end{equation}
According to~(\ref{eq: qy-series-2}) and~(\ref{eq: qy-series-4}), we write
\begin{equation} \label{eq: qy-series-5}
 \int_0^{+\infty}y^{-1}q^ \prime(y)dy = \sum_{\xi=1}^{+\infty}c_\xi\frac{\Gamma(a_\xi+1)} {b^{a_\xi+1}}   
\end{equation}
Using the expressions in~(\ref{eq: qy-series-3}) and~(\ref{eq: qy-series-5}), we obtain the equation~(\ref{eq:GS-4}) \qedsymbol.
\end{prf}

\subsection{Improved RKF-SL} \label{sec:RKF-SL}
Here, we give the MMSE estimator of the scale function under the RKF-SL framework. As a special case of the GSM distribution, it has been applied to the RKF in the previous work~\cite{RKF-GSM-1,RKF-SL-1}, whilst only the MAP estimate of the scale function was employed. To enhance the precision, we derived the MMSE estimate and Proposition~\ref{pro:slash estimation} is presented with details. 

\begin{pro} \label{pro:slash estimation}
 For a slash distribution with dof parameter $v$, $\mathrm{E}\left(\kappa(y)\right)$ can be estimated by, 
 \begin{equation}\label{eq:Slash-1}
\mathrm{E}\left(\kappa(y)\right)=\mathrm{E}\left(y\right) = \frac{\mathit{\gamma}(a+1,b)}{b\mathit{\gamma}(a,b)},  
\end{equation}
where
\begin{equation} \label{eq:Slash-2}
 a=\frac{m+v}{2}\quad \mathrm{and} \quad b=\frac{\eta}{2}.  
\end{equation}
\end{pro}

\begin{prf} \label{proof:slash estimation}
For a slash distribution, the mixing density follows $\mathrm{Be}(y;\frac{v}{2},1)$ and $0<y<1,v>0$. Then according to~(\ref{eq:logq_y}), we have 
\begin{equation*}
q^ \prime (y)\propto y^{\frac{m+v-2}{2}}\mathrm{exp}\left(-\frac{\eta}{2}y\right).   
\end{equation*}
Thus $q^ \prime(y)$ is proportional to a Gamma pdf when $0<y<1$. As $\kappa(y)=y$, we have
\begin{equation} \label{eq:qy-slash-1}
\mathrm{E}\left(\kappa(y)\right)=\mathrm{E}\left(y\right)=\frac{\int_0^{1}yq^ \prime(y)dy}{\int_0^{1}q^ \prime(y)dy}  
\end{equation} 
Calculate $a$ and $b$ according to~(\ref{eq:Slash-2}), then
\begin{equation} \label{eq:qy-slash-2}
\begin{split}
  \int_0^{1}yq^ \prime(y)dy &=\frac{\mathit{\gamma}(a+1,b)}{b^{a+1}} \\
   \int_0^{1}q^ \prime(y)dy &= \frac{\mathit{\gamma}(a,b)}{b^{a}}.  
\end{split}
\end{equation}
The expressions in~(\ref{eq:qy-slash-1}) and (\ref{eq:qy-slash-2}), give the expression in~(\ref{eq:Slash-1}) which finalises the proof \qedsymbol.
\end{prf}

\bibliographystyle{IEEEtran}
\bibliography{Bibliography}

\end{document}